\documentclass[twocolumn,preprintnumbers,amsmath,amssymb,aps,prb,reprint]{revtex4}
\usepackage{graphicx}
\begin{document}

\title{
Nonlinear Dynamics, Avalanches and Noise For Driven Wigner Crystals 
} 
\author{
C. Reichhardt and C. J. O. Reichhardt 
} 
\affiliation{
Theoretical Division and Center for Nonlinear Studies,
Los Alamos National Laboratory, Los Alamos, New Mexico 87545, USA\\ 
} 

\date{\today}
\begin{abstract}
  We consider the driven dynamics of Wigner crystals interacting with random disorder. Using numerical simulations, we find a rich variety of transport phenomena as a function of charge density, drive, and pinning strength. For weak pinning, the system forms a defect free crystal that depins elastically. When the pinning is stronger, a pinned glass phase appears that depins into a filamentary flow state, transitions at higher drives into a disordered flow phase, and finally forms a moving smectic. Within the filamentary flow phase, the conduction curves can show switching dynamics as well as negative differential conductivity in which switching events cause some flow channels to be blocked. The velocity noise in the filamentary flow regime exhibits narrow band characteristics due to the one-dimensional nature of the motion, while the moving smectic has narrow band velocity noise with a washboard frequency. In the disordered flow state, the noise power reaches a peak value and the noise has a $1/f$ character. Our transport results are consistent with recent experimental transport studies in systems where Wigner crystal states are believed to be occurring. Below the conduction threshold, we find that avalanches with a power law size distribution appear when there are sudden local rearrangements of charges from one pinned configuration to another.
\end{abstract}

\maketitle

\section{Introduction}
There is a broad class of particle like systems that exhibit
transitions from pinned to sliding states 
as function of driving when the particles are coupled to a
random substrate \cite{Reichhardt17}.
Some examples include depinning of vortices
in type-II superconductors \cite{Reichhardt17,Bhattacharya93}, 
colloidal particles
driven over rough landscapes \cite{Reichhardt02,Pertsinidis08},
frictional systems \cite{Vanossi13},
magnetic skyrmions \cite{Reichhardt15,Jiang17},
and active matter \cite{Sandor17a,Morin17}.
In these systems, there is a critical threshold force
that must be applied in order
for the particles to remain in
continuous motion.
If the quenched disorder is weak, the depinning transition can be
elastic, with all of the particles maintaining the same neighbors. 
In contrast, for strong quenched disorder, the 
depinning becomes plastic
and the system breaks up or phase separates into
moving and pinned regions \cite{Reichhardt17}. 
Even when the quenched disorder is very strong, at
higher drives it is possible for a 
dynamic reordering transition
to occur into a moving anisotropic crystal
\cite{Reichhardt17,Giamarchi96} or moving smectic
state \cite{Giamarchi96,Balents98,Olson98a}. 
The shape of the velocity-force curves near
depinning varies depending on whether the depinning
is elastic or plastic or if thermal creep comes into play
\cite{Reichhardt17}.
In some cases, the depinning curves are
smooth and have the form
$V \propto (F_{D} - F_{c})^\beta$ \cite{Reichhardt17,Fisher98,Fily10},
where $V$ is the velocity, $F_D$ is the driving force, and $F_c$ is the
depinning threshold.
In a two-dimensional (2D) system,
the elastic depinning exponent $\beta = 2/3$, 
while for plastic depinning, $1.25 < \beta < 2.0$. 
At higher drives, one or more peaks can appear in
the $d\langle V\rangle/dF_{D}$ curves, serving as signatures of 
different dynamical transitions. 

In the plastic flow regime, an open question is whether
there can be different types of
coexisting flow states,
such as plastic flow that is fluctuating or plastic flow
that forms specific stable structures.
If so, this would imply that additional transitions are possible
between different types of plastic flow.
For example, one form of plastic flow could be
dominated by a finite number of filamentary flow channels.
If the filamentary structures change,
the velocity-force curves would not be smooth but would
show switching behavior with
a series of jumps corresponding to the drives at which
different channels of motion open for flow
\cite{Reichhardt17,GronbechJensen96,Reichhardt98}.
Switching effects and negative differential conductivity occur in 
systems
such as sliding charge density waves \cite{Gruner88},
superconducting vortices in mesoscopic channels \cite{Reichhardt98,Besseling05}
or near the peak effect \cite{Bag17},
and electron liquid crystals \cite{Cooper03,Qian17};
however, it is not well understood how these effects
are connected to plastic flow,
which often produces nonlinear but smooth velocity-force curves.

Depinning and different sliding phases
can also be explored by measuring noise
fluctuations, which can show distinctive
features such as broad band noise
in strongly plastic regimes \cite{Reichhardt17,Olson98a,Marley95}
or narrow band noise
\cite{Reichhardt17,Olson98a,Kolton99,Togawa00,Okuma07}
when the system forms a moving crystal.
Other studies have shown that
transitions from ordered to disordered
states are associated with a shift
from narrow to broad band noise
signatures \cite{Cooper03,Diaz17,Sato19,Sun22}.

Another example of systems that
can exhibit a variety of threshold and switching behaviors
is Wigner crystals
\cite{Andrei88,Goldman90,Williams91,Jiang91,Zhu94,Cha94,Reichhardt01,Monceau12,Brussarski18,Yoon99,Hossain22,Rees16,Badrutdinov16,Lin18}
and electron liquid crystals
\cite{Cooper03,Qian17,Sun22,Fogler96,Moessner96,Fradkin99,Lilly99a,Cooper99,Reichhardt03a,Wang15}.
Wigner crystals can arise
in a number of 2D systems, and the appearance of a threshold for
conduction has been interpreted
as the depinning of Wigner crystals.
More recently,
a study of the sliding of Wigner crystals
or electron solids for varied carrier concentrations
in Ref.~\cite{Brussarski18} showed that
the depinning threshold decreases
with increasing carrier density, and that
there can be a two step depinning process
followed by a cusp above which the
current increases linearly with increasing drive.
In this same work, there was
a peak in the noise power near depinning as
well as $1/f$ noise characteristics.

Evidence for Wigner crystal formation has been
found with numerous resonance experiments
\cite{Li00,Ye02,Chen04,Chen06,Jang17},
magnetoresistance measurements \cite{Zhang15,Knighton18,Hossain20,Shingla21},
and commensurability oscillations for composite fermions encircling
charges in a Wigner crystal \cite{Jo18}. Several
experiments have 
shown finite depinning thresholds accompanied by
negative differential resistance in which
the conduction drops with increasing
drive \cite{Csathy07,Madathil22}.
There has also been work 
on 2D metal-insulator transitions suggesting that
the system could be forming a Wigner crystal
\cite{Jaroszynski02,Leturcq03,Jaroszynski04}
or a mixed Wigner crystal and fluid state
\cite{Abrahams01,Spivak10}, and the noise
behavior is consistent with what
is seen for models of transitions from Wigner crystals
to Wigner glasses in the presence of quenched disorder \cite{Reichhardt04}. 

More recently, evidence
for Wigner crystals has been found in monolayer semiconductors
and dichalcogenide monolayers \cite{Smolenski21},
moir{\' e} heterostructures \cite{Li21}, and oxides \cite{Padhi18},
while evidence for bilayer Wigner crystals
appears in dichalcogenide
heterostructures \cite{Zhou21}.
There have also been
several predictions for Wigner crystal formation in the insulating states
of moir{\' e} systems \cite{Padhi18,Padhi21}.
New methods have recently been developed for creating high quality
2D election systems that should allow for easier
access to Wigner crystal states \cite{Chung21}.
A major question is what is
the role of quenched disorder in
Wigner crystal phases and dynamics,
and how it can be determined if a
system is in a Wigner crystal, Wigner glass, or fluid state
\cite{Shayegan22}.

Given the recent evidence for Wigner crystal formation and
the ubiquity of disorder in most of these systems, it
is interesting to study in detail the
different possible threshold behaviors and
charge flow patterns for a Wigner crystal
coupled to disorder
to see if there are distinct phases,
or whether effects such
as switching, negative differential conductivity,
and transitions among narrow band and broad band noise characteristics
can occur.

Here we examine the driven dynamics of a Wigner
crystal interacting with random quenched
disorder using molecular dynamics simulations.
Previous work employing this approach focused on the depinning threshold and
how it changes as a transition
occurs from an ordered crystal to a defected solid
as a function of increasing quenched
disorder strength \cite{Cha94,Cha94a}.  Other
work focused on the dynamic ordering
transition from a plastically flowing crystal
into a dynamically ordered moving
crystal or smectic state for
increasing drive \cite{Reichhardt01},
similar to the dynamic ordering
found in driven
superconducting vortex lattices \cite{Giamarchi96,Balents98,Olson98a}.
In additional studies,
the change in the noise fluctuations across a pinned Wigner
glass to fluid state transition was examined \cite{Reichhardt04}.
More recently,
molecular dynamics 
simulations were used to explore the impact of the magnetic field on the
sliding dynamics of Wigner crystals,
and it was shown that the system has
a velocity-dependent Hall angle due to a side jump effect
of the charges moving over
the pinning sites \cite{Reichhardt21},
similar to the
velocity-dependent Hall angle
observed
in
magnetic skyrmion systems with pinning \cite{Reichhardt15,Jiang17,Reichhardt22}.

In this work we focus on the Wigner crystal dynamics
for both elastic and plastic depinning.
When the quenched disorder is strong, 
a disordered state appears that
can depin into a filamentary flow phase where the motion occurs
in quasi-one-dimensional (q1D) rivers or channels.
The opening of these channels leads to jumps
in the conduction, and in some cases,
the motion decreases with increasing drive due to the closing of
a flow channel,
leading to negative differential conductivity.
We find that the conduction threshold deceases with increasing
carrier density as a result of the appearance of interstitially
pinned charges that are immobilized only by the repulsion from other
charges and not directly by a pinning site. This occurs
once the charge density is high enough that the strongest
pinning sites are fully occupied by charges.
When a sudden drive is applied
in the filamentary
flow regime,
a strongly fluctuating transient state can appear
before the system settles into well defined flowing
filaments with a low frequency narrow band noise signal.
At higher drives, the filamentary flow
is replaced by a
strongly fluctuating state in which the noise has a
low frequency $1/f^{\alpha}$ character with
an exponent $\alpha = 0.7$ that is close to the value found
in recent experiments \cite{Brussarski18}.
At higher frequencies, the noise signature has a
$1/f^2$ form.
The overall
noise power at a specific frequency shows
a strong peak just above the conduction
threshold,
similar to what is observed in recent experiments \cite{Brussarski18}.
For
even higher drives, the noise power is reduced
and narrow band noise appears
when the
system enters a moving smectic or moving crystal state.
In the strong pinning regime there can be a three step depinning
process in which filamentary flow is followed by a nonlinear regime
and then by a cusp above which
the velocity increases linearly with drive, similar to
transport curves obtained in recent experiments \cite{Brussarski18}.

In the pinned state,
as the drive is increased below the threshold for plastic flow,
there can be sudden bursts of motion or avalanches
in the form of q1D flowing channels of different lengths.
The sizes $s$ of these avalanches are power law distributed
according to $P(s) \propto s^{-1.7}$.
In the elastic pinning regime, the system forms
a crystal and depins without the generation of topological defects,
and there is no switching behavior or filamentary flow.
In the plastic depinning regime,
the Wigner crystal exhibits
a narrow band noise signal just above the depinning threshold.

\section{Simulation and System} 

We consider a 2D system
of size $L \times L$ with periodic boundary conditions
in the $x$ and $y$ directions
containing $N$ classical electrons or charges that couple
to quenched disorder.
The charge density is $n=N/L^2$ and we set $L=36$.
After obtaining an initial configuration via simulated annealing,
we increase the driving force from zero in small increments applied over
fixed windows of time in order to observe any transient motion that occurs.
The first drive
that is strong enough to produce continuous motion is defined to be
the conduction threshold.

The equation of motion for charge $i$ in the Wigner crystal is
\begin{equation} 
\alpha_d {\bf v}_{i} = \sum^{N}_{j \neq i}\nabla U(r_{ij}) +  {\bf F}_{p} + {\bf F}_{D} \ ,
\end{equation}
where $\alpha_{d}$ is the damping constant.
The charge-charge interaction potential
$U = e^2/r_{ij}$, where $r_{ij}=|{\bf r}_i-{\bf r}_j|$
is the distance between electrons located at ${\bf r}_i$ and ${\bf r}_j$,
and $e$ is the electronic charge.
The pinning force ${\bf F}_{p}$ 
is modeled as arising from $N_p$ short range parabolic traps
of finite radius $r_p$ that exert a maximum confining force
of $F_p$ on a charge, while
the driving force ${\bf F}_D=F_D{\bf \hat x}$ models an externally applied
electric field.
The pinning density $n_p=N_p/L^2$ is fixed to $n_p=0.25$.
We define the filling fraction $F=N/N_p$ to be the ratio of the number of
charges to the number of pinning sites.
Since the charge-charge interactions
are long range, we employ a Lekner summation technique
for computational efficiency as used in previous studies of particles
with Coulomb interactions \cite{Reichhardt01,Reichhardt04,Reichhardt21}.
We also note that if the charges are in a magnetic field,
their motion is subjected to a Hall angle produced by
the Magnus force $e{\bf B}\times {\bf v}_{i}$.
In general, this
term is small, and
in previous work it was shown that the effect of the Magnus
force is particularly weak near the conduction threshold \cite{Reichhardt21}.
Since we focus specifically on the conduction
threshold in this work, we neglect the Magnus force.

We consider a range
of pinning forces, charge densities, and drive values, and employ
two types of driving protocols.
In the first, the drive is continuously
increased by
small increments to mimic
an experimental velocity-force curve
in which a current is swept from 
zero
up to a final value.
In the second,
we apply a constant drive and hold the drive
fixed for an extended period of time.
The latter approach allows us to obtain long time series
for examining conduction noise.
For the continuous sweep protocol,
we increase the external drive in increments of
$\Delta F_{D} = 0.0001$ and wait 18400 simulation time steps before
applying the next increase.
At each drive increment we measure the
average velocity per charge in the driving direction,
$\langle V_x\rangle = N^{-1}\sum^{N}_i{\bf v_i}\cdot {\hat {\bf x}}$.
We also calculate the average perpendicular velocity,
$\langle V_y\rangle = N^{-1}\sum^{N}_i{\bf v_i}\cdot {\hat {\bf y}}$,
as well as the 
standard deviations of the velocities
for both directions,
$\delta V_{x} = \sqrt{[\sum^{N}_{i}({\bf v_i}\cdot {\hat {\bf x}})^2
    - \langle V_{x}\rangle^2]/N}$ and
$\delta V_{y} = \sqrt{[\sum^{N}_{i}({\bf v_i}\cdot {\hat {\bf y}})^2 - \langle V_{y}\rangle^2]/N}$. 
In certain regimes, we increase the waiting time
in order to obtain
smoother curves for calculating d$\langle V_x\rangle/dF_D$.
This increase also allows us to clearly
delineate bursts or avalanches of motion.

\section{Switching and Conduction}

\begin{figure}
\includegraphics[width=\columnwidth]{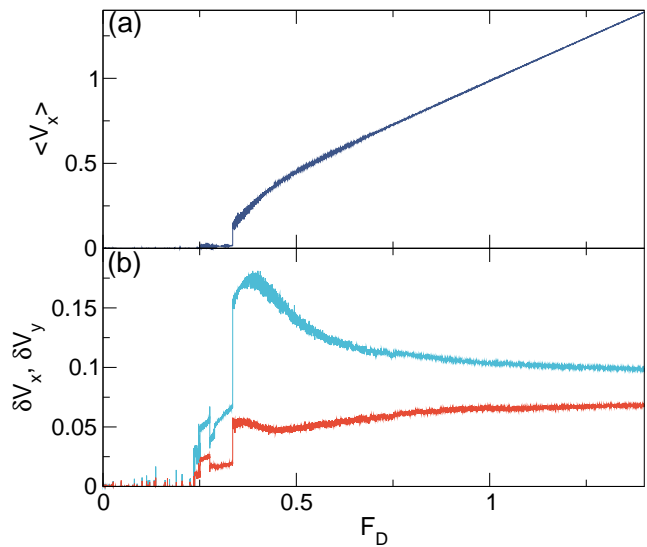}
\caption{
(a) The velocity $\langle V_x\rangle$
in the driving direction vs the driving force $F_{D}$ for a system
with a filling factor of $F=0.4$ and
a pinning force of $F_{p} = 0.5$.
The onset of conduction occurs
near $F_{D} = 0.35$.
(b) The corresponding standard deviations of the velocity
parallel, $\delta V_x$ (blue), and perpendicular, $\delta V_y$ (red),
to the driving direction vs $F_D$ for the same system, 
showing
a pinned regime and a filamentary flow regime
for drives below the jump into the disordered flow regime.
}
\label{fig:1}
\end{figure} 

In Fig.~\ref{fig:1}(a) we plot the velocity $\langle V_x\rangle$
versus the driving force $F_{D}$ for
a system with a pinning strength of $F_{p} = 0.5$
and a filling fraction of $F = 0.4$ under a continuously increasing drive.
A clear conduction threshold appears near
$F_{D} = 0.34$.
In Fig.~\ref{fig:1}(b) we plot the corresponding standard deviations
of the velocities parallel, $\delta V_x$, and perpendicular, $\delta V_y$,
to the drive versus $F_D$.
These curves show more clearly that below the conduction threshold,
for $0.24 < F_{D} < 0.34$,
there are a series of jumps in the velocity fluctuations in both the $x$
and $y$ directions, while
at $F_{D} = 0.34$ there is a large peak in $\delta V_{x}$ at the
transition to the
plastic or continuously fluctuating
disordered flow phase.
In general, $\delta V_{y} < \delta V_{x}$.  
For $F_{D} > 0.65$,
there is an
ordering into a moving smectic phase. 
The onset of this transition extends down to
$F_{D} > 0.4$, where $\delta V_{x}$ begins to drop as the
flow gradually becomes less disordered and the smectic
state emerges.

\begin{figure}
\includegraphics[width=\columnwidth]{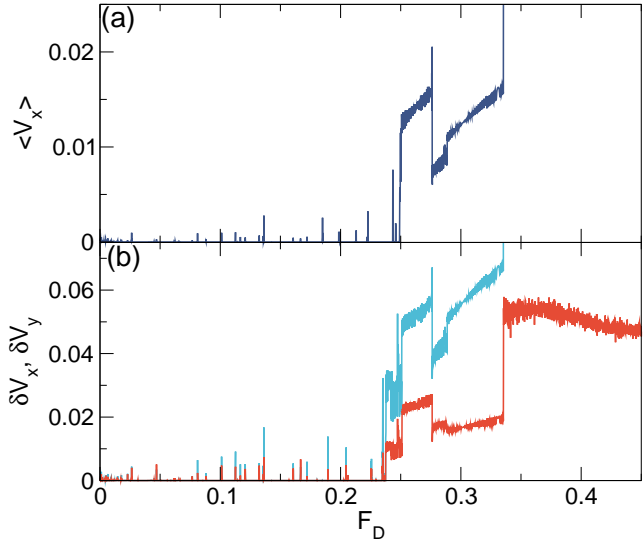}
\caption{
Blow ups of  
(a) $\langle V_x\rangle$ vs $F_D$ and
(b) $\delta V_x$ (blue) and $\delta V_y$ (red) vs $F_D$
near the conduction threshold for the system in Fig.~\ref{fig:1} with
$F=0.4$ and $F_p=0.5$.
There are finite jumps both up and down in all three quantities.
There are also a number of isolated jumps within the
pinned phase that correspond to avalanche motion.
}
\label{fig:2}
\end{figure}

In Fig.~\ref{fig:2}(a,b)
we show a closeup of  $\langle V_{x}\rangle$,
$\delta V_{x}$, and $\delta V_y$ versus $F_D$
for the system from Fig.~\ref{fig:1}.
There is a pinned regime
for $F_{D} < 0.24$ in which $\langle V_x\rangle = 0$,
followed by what we call the filamentary flow phase
for $0.24 < F_{D} < 0.34$.
We note that even within the pinned phase,
there are still a number of isolated jumps
in all three quantities that
correspond to avalanches involving the
correlated motion of multiple charges when the drive is increased.
At $F_{D} = 0.25$,
there is a large jump in $\langle V_x\rangle$ followed by a region in
which $\langle V_x\rangle$ increases linearly with increasing $F_D$.
This is followed by a jump down
in $\langle V_x\rangle$ and then
by a smaller and a larger jump up.
Counterparts to these jumps
also appear in $\delta V_{x}$ and $\delta V_y$.
The upward jumps are associated with switching events
in which
well defined channels of motion appear,
while downward jumps occur
when the channels close or change shape and the
flow is reduced.
Once a channel has formed, the velocity of the charges in
that channel increases with increasing $F_D$
until there is a sudden rearrangement of charges somewhere in the system
that 
creates
a new channel or shuts down one or more of the existing channels.
The jump down in $\langle V_{x}\rangle$ for increasing
$F_{D}$ is indicative of negative differential conductivity.
For $F_{D} \geq 0.34$, the
channels become more chaotic and
randomly change over time, causing the overall flow
to be more two-dimensional in character.

\begin{figure}
\includegraphics[width=\columnwidth]{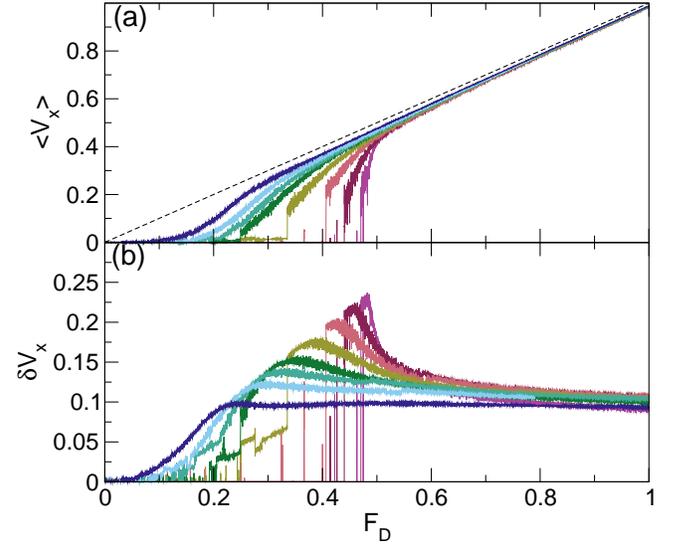}
\caption{
(a) $\langle V_x\rangle$ versus $F_{D}$ for the system from
Fig.~\ref{fig:1} with $F_{p} = 0.5$ at fillings of
$F = 1.4$ (dark blue), 1.0, 0.8, 0.6, 0.4, 0.3, 0.2, and 0.1 (light magenta),
from left to right. The dashed line indicates the
velocity response in a sample with no pinning.
(b) The corresponding $\delta V_{x}$ versus $F_D$.
The depinning threshold decreases with increasing
charge density,
and the filamentary phase only occurs for $F > 0.3$.
}
\label{fig:3}
\end{figure}

In Fig.~\ref{fig:3}(a) we plot $\langle V_{x}\rangle$
versus $F_{D}$ for the system from Fig.~\ref{fig:1} with
$F_p=0.5$ at different
filling factors of $F = 1.4$, 1.0, 0.8, 0.6, 0.4, 0.3, 0.2, and $0.1$,
while
Fig.~\ref{fig:3}(b) shows the corresponding $\delta V_{x}$ versus $F_D$.
As the filling factor, and thus the charge density, increases,
the depinning threshold shifts to lower drives,
in agreement with experimental observations \cite{Brussarski18}.
The dashed line in Fig.~\ref{fig:3}(a) indicates the velocity response
that would appear in a system with no pinning.
The jump in the velocity response at the
transition to continuous disordered
flow is sharper for lower fillings.
There is a peak in $\delta V_{x}$ at the onset of the disordered
flow regime, followed by a decrease in $\delta V_x$ with increasing $F_D$.
The $\delta V_x$ curves indicate that
the filamentary flow highlighted in
Fig.~\ref{fig:2} only occurs for $F > 0.3$.

\begin{figure}
\includegraphics[width=\columnwidth]{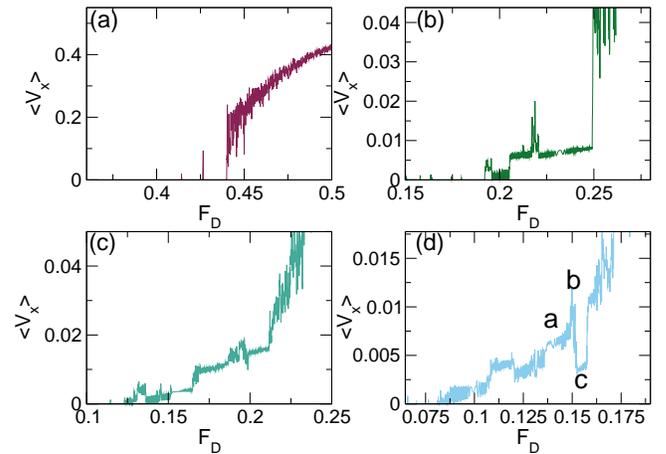}
\caption{A blowup of the
$\langle V_x\rangle$ vs $F_D$ curves for
selected samples from Fig.~\ref{fig:3}
with $F_p=0.5$.  
(a) $F = 0.2$ where there is no filamentary regime.
(b) $F = 0.6$ showing a filamentary regime.
(c) $F = 0.8$.
(d) $F = 1.0$. The letters a, b, c in panel (d)
indicate the values of $F_{D}$ at which the images in
Fig.~\ref{fig:5}(a,b,c) were obtained.}
\label{fig:4}
\end{figure}

In Fig.~\ref{fig:4}(a) we show a closeup of the $\langle V_x\rangle$
versus $F_D$ curve
for the system in Fig.~\ref{fig:3}
at $F = 0.2$, where a single depinning transition
occurs at $F_{D} = 0.44$ from
a pinned phase to
a disordered or plastic flow phase, and
there is no filamentary flow.
Figure~\ref{fig:4}(b) shows the same
system at $F = 0.6$, which depins near $F_D=0.19$ into a
filamentary flow phase and then exhibits
several upward and downward jumps in the velocity.
Immediately after most of these jumps,
the velocity increases linearly with $F_D$ and shows an oscillatory
behavior that we discuss in section IV.
Near $F_D = 0.25$ there is a transition to
the plastic fluctuating flow phase.
In Fig.~\ref{fig:4}(c), a sample with
$F = 0.8$ has an extended filamentary regime,
while in Fig.~\ref{fig:4}(d),
at $F = 1.0$, there is a filamentary regime
showing numerous switches corresponding to
both positive
and negative differential conductivity.
As Fig.~\ref{fig:4} illustrates,
the filamentary flow regime is robust over a range of fillings,
and in general the extent of the filamentary flow increases as
$F$ becomes larger.

\begin{figure}
\includegraphics[width=\columnwidth]{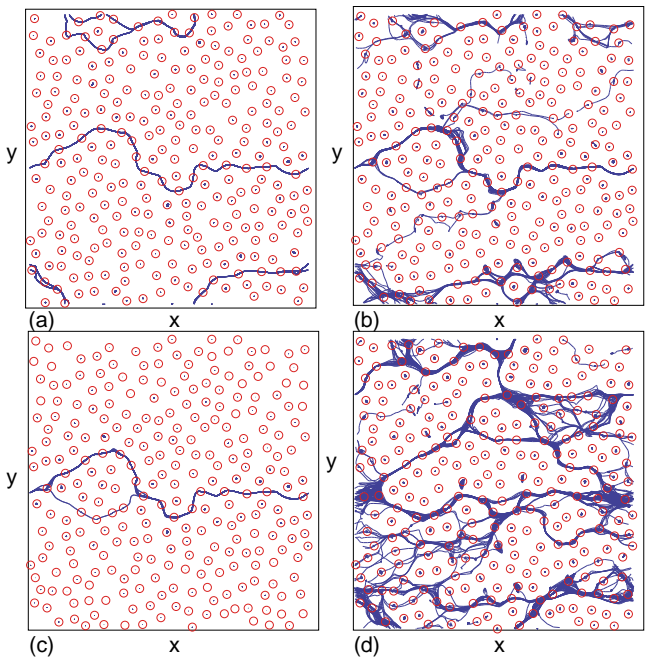}
\caption{
Images of the charge locations (red circles)
and trajectories (lines) for the
system in Fig.~\ref{fig:4}(d) with $F_p=0.5$ and
$F=1.0$. The letters in Fig.~\ref{fig:4}(d) correspond to the
values of $F_{D}$ at which the images were obtained.
(a) Stable filamentary channels at $F_D=0.137$.
(b) The filamentary regime just after a switching event at $F_D=0.15$
where the number of flow channels increases, $\langle V_x\rangle$ 
jumps upward, and the channels gradually change over time.
(c) The filamentary regime at $F_D=0.152$ after a switch in which
$\langle V_x \rangle$ drops and the flow is confined to a single
branched channel.
(d) Fluctuating channels
at $F_{D} = 0.2$.
}
\label{fig:5}
\end{figure}

\begin{figure}
\includegraphics[width=\columnwidth]{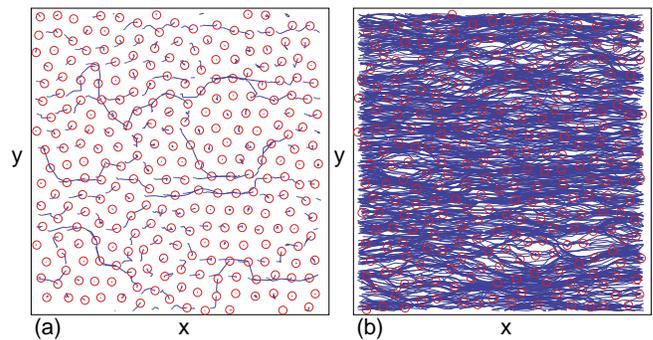}
\caption{
Images of the charge locations (red circles) and trajectories (lines) for the
system in Fig.~\ref{fig:4}(d) with $F_p=0.5$ and $F=1.0$
in the disordered regime at $F_{D} = 0.3$.
(a) A short time trajectory image shows that the flow is 2D in character
and that moving
and temporarily pinned charges coexist.
(b) A longer time trajectory image indicates that all of the charges
participate in the motion over time and the behavior is liquid like.
}
\label{fig:6}
\end{figure}

To get a better picture of the flow in the
different phases,
in Fig.~\ref{fig:5} we show the charge trajectories for the drives
labeled with letters in
Fig.~\ref{fig:4}(d).
At $F_D=0.137$ in Fig.~\ref{fig:5}(a),
there are two q1D winding flow channels or filaments. As $F_D$
increases,
the same channels persist but the flow through each channel
becomes faster.
At $F_D=0.15$,
there is a sudden rearrangement
of the charges
leading to a jump up
in $\langle V_x\rangle$ since there are now more channels flowing,
as illustrated in Fig.~\ref{fig:5}(b).
The channels
of Fig.~\ref{fig:5}(b) slowly change over time, whereas
the channel structure in Fig.~\ref{fig:5}(a) is static.
At $F_D=0.152$, there is a transition to
the single stable channel state shown in
Fig.~\ref{fig:5}(c),
and $\langle V_x\rangle$ drops below the value it had
in Figs.~\ref{fig:5}(a) and (b) even though $F_D$ is higher.
This channel structure remains stable under increasing
$F_{D}$
until the next switching event
occurs.
In some switching events, there is a transition to
fluctuating channels, while in other events,
a transient fluctuating state settles into a stable filamentary flow
state.
The general features found in
Fig.~\ref{fig:5}(a,b,c) also occur
for the other jumps in $\langle V_x\rangle$
in the filamentary flow phase of Fig.~\ref{fig:4}(d).
As $F_{D}$ is further
increased, the channel structure eventually breaks down and is replaced
with a disordered and continuously changing flow, such as that shown
in Fig.~\ref{fig:5}(d)
at $F_{D} = 0.2$.
At even higher drives, the disordered flow becomes more two-dimensional
in character and consists of coexisting moving and temporarily
pinned charges, as illustrated in Fig.~\ref{fig:6} at $F_D=0.3$.
The short time trajectories in Fig.~\ref{fig:6}(a)
show that there is a coexistence of moving and temporarily pinned
charges, while the longer time trajectories of
Fig.~\ref{fig:6}(b) indicate that there are no permanently pinned regions
and the overall flow is liquid like in nature.

Flow through q1D channels
has also been observed for other
systems of particles moving over quenched disorder
\cite{Reichhardt17};
however,
in most of these systems
there is not a clearly defined filamentary flow phase.
Instead, systems with shorter range interactions show
a continuous crossover between filamentary flow
and a 2D disordered flow phase.
Gr{\o}nbech-Jensen {\it et al.} \cite{GronbechJensen96} considered
a 2D simulation of vortices in thin film superconductors
and found a clear region of filamentary flow at lower drives
followed by disordered 2D flow at higher drives.
The interactions between vortices in thin film superconductors obey
a $\ln(r)$ potential or a $1/r$ force,
in contrast to vortex lines in bulk superconducting crystals, where
the interaction is a Bessel function that is similar to a 
screened Coulomb potential.
Gr{\o}nbech-Jensen {\it et al.}
argued that due to the long range interactions in the thin film
system,
the shear modulus $C_{66}$ of the vortex lattice
is much lower than the compression
modulus,
so the vortices can easily slide past each other, favoring filamentary
motion.
The filamentary flow can also be viewed as a consequence of the existence
of interstitial vortices, which occupy the spaces between pinning sites and
are immobilized only due to the repulsion of
neighboring pinned vortices rather than being trapped directly by a pinning
site.
If the interaction potential between vortices is very smooth, as is the
case for a $\ln(r)$ potential,
the potential landscape experienced by an interstitial vortex will be
fairly flat.
For shorter range
interaction potentials,
the landscape looks more like a series of obstacles in which the
vortices can become trapped.
This suggests that for a Wigner crystal with no screening,
where the charge-charge interactions are of long range,
filamentary flow phases should be a general feature.
On the other hand,
if screening is occurring,
the behavior will be closer to that of bulk superconducting vortices or
particles with short range interactions,
and the filamentary flow phase will be lost.

\begin{figure}
\includegraphics[width=\columnwidth]{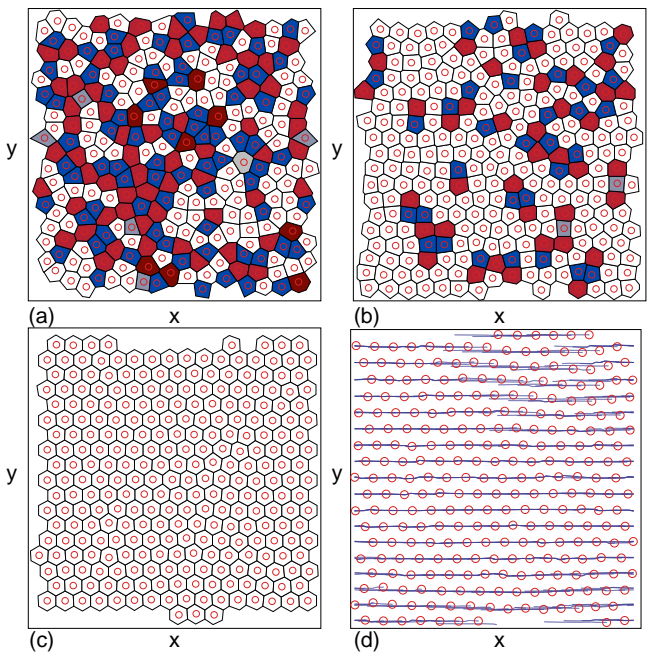}
\caption{
(a,b,c) Voronoi constructions of the charge locations (red circles)
in the system
from Fig.~\ref{fig:4}(d) with $F_p=0.5$ and $F=1.0$.
Polygon colors indicate the coordination number of each electron:
4 (dark grey), 5 (blue), 6 (white), 7 (light red), 8 (dark red), 9 (light grey).
(a) $F_{D} = 0.25$.
(b) $F_{D} = 0.45$.
(c) $F_{D} = 0.8$.
The system becomes more ordered with
increasing drive and forms a moving smectic at the highest drive.
(d) Image of the charge locations (red circles) and trajectories (lines)
for the sample in panel (c) with $F_D=0.8$,
showing straight non-crossing channels of flow.}
\label{fig:7}
\end{figure}

At higher drives, the charges dynamically order into a moving crystal or
moving smectic state.
In general, there are still a small number of lattice defects
present, so the
system is best described as a weak moving smectic.
In Fig.~\ref{fig:7}(a,b,c) we show Voronoi constructions of the charge
positions at $F_D=0.25$, 0.45, and 0.8, respectively, where polygon colors
indicate the coordination number $z_i$ of each charge.
The system is strongly disordered
at $F_{D} = 0.25$ in Fig.~\ref{fig:7}(a) and contains numerous topological
defects,
while at $F_{D} = 0.45$ in Fig.~\ref{fig:7}(b), the number
of defects is reduced.
In Fig.~\ref{fig:7}(c) at $F_{D}= 0.8$, the
lattice is almost completely triangular
and contains only sixfold coordinated
charges. The charge trajectories at $F_D=0.8$, shown in
Fig.~\ref{fig:7}(d), follow straight q1D channels that do not intersect.
Dynamical reordering transitions for driven Wigner crystals
was studied previously in a system
with long range pinning sites \cite{Reichhardt01}.
In the present study, the pinning sites are of short range, 
and the Wigner crystal is able to organize into a state containing almost
no topological defects. In contrast, in Ref.~\cite{Reichhardt01}, the
long range pinning interfered with the dynamical reordering and the system
only reached a 
moving smectic
state in which a finite number of topological defects persist that are
aligned so as to glide along the driving direction. 

\begin{figure}
\includegraphics[width=\columnwidth]{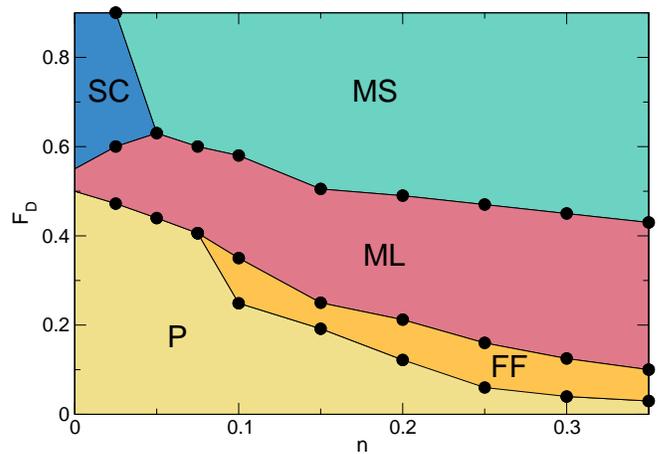}
\caption{
Dynamic phase diagram as a function of
driving force $F_{D}$ vs
charge density $n$ for the system in
Figs.~\ref{fig:4} and \ref{fig:5} with $F_{p} = 0.5$.
Since the pinning density is always fixed to $n_p=0.25$,
the filling factor $F=N/N_p=n/n_p=4n$.
The phases are: P, pinned (yellow);
FF, filamentary flow (orange);
ML, disordered moving liquid (pink);
SC, moving semi-crystallized state (blue);
MS, moving smectic (green).
}
\label{fig:8}
\end{figure}

From the features in the
$\langle V_x\rangle$,
$\delta V_{x}$, and $F_{D}$ curves,
along with the images of the topological defects,
we construct a dynamic phase diagram
as a function of $F_{D}$ versus
charge density $n$
for a system with fixed $F_{p} = 0.5$, as shown in Fig.~\ref{fig:8}.
Since the pinning density is always fixed at $n_p=0.25$,
we have $F=N/N_p=n/n_p=4n$ and $n=F/4$.
For $n < 0.1$, the system depins directly into a disordered
moving liquid phase and the filamentary flow state is absent,
as shown in Fig.~\ref{fig:4}(a).
At high drives, a moving smectic forms if the charge density is high enough,
and in some cases the system is small enough for this smectic to
completely order into a moving crystal without defects.
When the charge density is very low,
some randomly oriented topological defects remain present
even at high drives and
the system forms a partially ordered crystal or semi-crystal.
For $n > 0.075$, a
window of filamentary flow phase appears above depinning,
as illustrated in
Fig.~\ref{fig:5}(a,b,c),
followed by a transition into the fluctuating disordered moving liquid.
The filamentary flow phase is lost
at low charge densities because all of the charges
are able to occupy strong pinning sites and there are no
interstitial charges.
The depinning threshold marking the end of the pinned phase drops
to lower $F_D$ with increasing $n$,
in agreement with
recent experiments \cite{Brussarski18}.

\section{Noise Measures}

\begin{figure}
\includegraphics[width=\columnwidth]{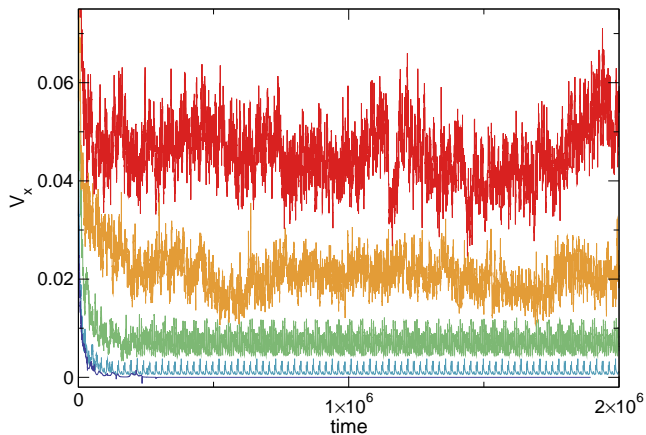}
\caption{
Time series of the velocity $V_x$ after application of a pulse drive  
for a system with $F = 1.0$ and $F_{p} = 0.5$.
At $F_{D} = 0.04$ (dark blue), the system reaches a pinned state.
For $F_D=0.075$ (light blue) and $F_D=0.1375$ (green), the system settles into
filamentary flow 
with a periodic velocity signal.
For $F_{D} = 0.175$ (orange) and $F_D=0.2$ (red), the
system settles into a fluctuating state.
}	
\label{fig:9}
\end{figure}

We next consider the
system from Fig.~\ref{fig:4}(d) with
$F_p=0.5$ and $F=1.0$ but for a pulsed drive, where
we apply a fixed $F_{D}$
and wait for the system to settle
into a steady state before
measuring the noise fluctuations and obtaining the
long time average velocity.
In Fig.~\ref{fig:9} we plot $V_x$ versus time in
simulation time steps
for pulse drive amplitudes of $F_D=0.04$, 0.075, 0.1375, 0.175, and 0.2.
For $F_{D} = 0.04$, there is some initial transient motion, but
at longer times all of the charges become pinned
and the velocity drops to zero.
At $F_{D} = 0.075$ and $F_{D} = 0.1375$,
after an initial decreasing transient,
the velocities settle into a periodic pattern reflecting the formation of
a filamentary flow state containing one or more flow channels of
repetitive motion.
Unlike the simulations of section III
in which the drive is continuously increasing,
for the pulsed drive the filamentary channels adopt a stable configuration
and there is no switching.
In general, for the filamentary flow the system
settles into a state with a periodic velocity signal.
The periodicity of the signal can be
fairly complicated if there are multiple
filaments of flow present that
generate multiple frequencies simultaneously.
For $F_{D} = 0.175$ and $F_D=0.2$, there is still
an initial transient decay of the velocity but
the system remains in a fluctuating state
at long times.

\begin{figure}
\includegraphics[width=\columnwidth]{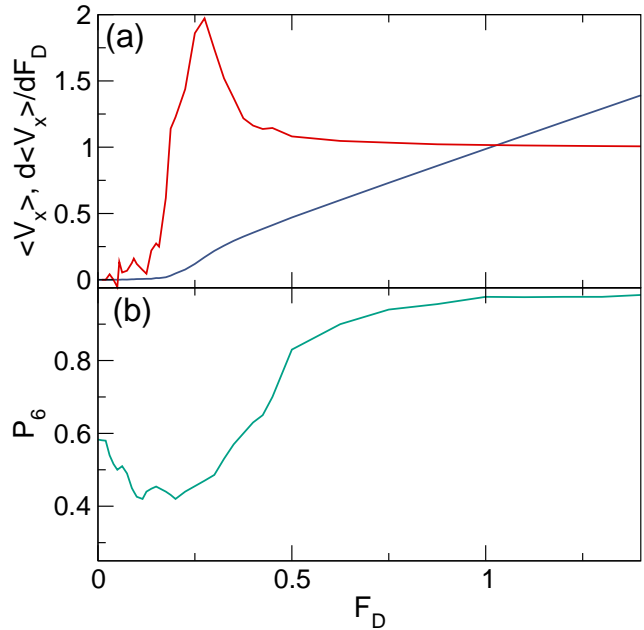}
\caption{
(a) The average velocity $\langle V_x\rangle$ (blue) and
its derivative $d\langle V_x\rangle/dF_D$ (red) vs $F_D$  
for the system in Fig.~\ref{fig:9} with $F=1.0$, $F_p=0.5$, and
pulse driving,
showing a two step depinning process.
(b) The corresponding
average fraction of sixfold coordinated
charges $P_{6}$ vs $F_{D}$.
The system is the most disordered in the
fluctuating phase and undergoes a dynamical
reordering transition at higher $F_{D}$.
}
\label{fig:10}
\end{figure}

By performing a series
of pulse measurements, we obtain
the average
velocity $\langle V_x\rangle$ and
its derivative $d\langle V_x\rangle/dF_{D}$ as a function of
$F_D$, as plotted in
Fig.~\ref{fig:10}(a) for the system
in Fig.~\ref{fig:9}.
The initial depinning
into a filamentary flow state occurs near $F_{D} = 0.0625$,
and the filamentary flow,
which extends from $0.0625 \leq F_{D} < 0.175$, produces a
series of small
jumps in $d\langle V_x\rangle/dF_{D}$.
For $0.175 \leq F_D < 0.5$, the
system enters a disordered strongly fluctuating
flow phase associated with
a large peak in $d\langle V_x\rangle/dF_D$, while for
$F_{D}\geq 0.5$, the velocity increases linearly with $F_D$ and
$d\langle V_x\rangle/dF_{D}$ approaches one.
The behavior of the velocity-force curve is similar to
recent experimental observations of nonlinear
velocity-force signatures
showing a two step depinning process into a nonlinear regime
at the onset of conduction followed by a transition to a linear
regime at higher
drives \cite{Brussarski18}.
We show in section V that as the strength $F_p$ of the disorder increases,
the
multiple step depinning
transitions become even more prominent.
Figure~\ref{fig:10}(b) illustrates the
fraction of the
average number of sixfold coordinated particles,
$P_{6}=N^{-1}\sum_i^N \delta(z_i-6)$,
versus $F_{D}$.
A local dip in $P_6$ occurs at the transition
from filamentary flow to fluctuating
flow, and there is a large increase
in $P_6$ near $F_{D} = 0.5$ when
the system transitions from the plastic disordered flow state
to a moving smectic.

\begin{figure}
\includegraphics[width=\columnwidth]{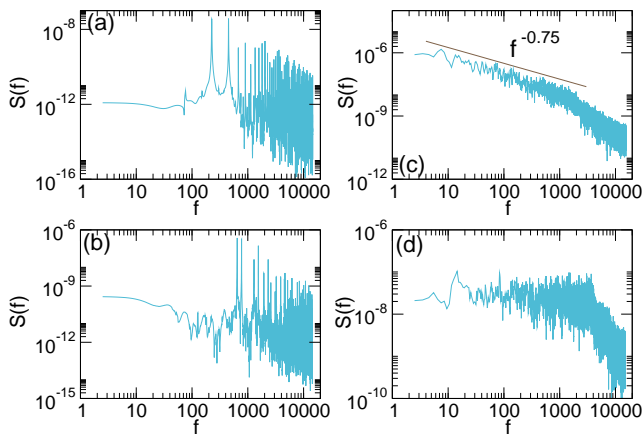}
\caption{
Velocity noise power spectra $S(f)$ for the pulse drive
system from Figs.~\ref{fig:9} and \ref{fig:10} with $F=1.0$ and
$F_p=0.5$.
(a) At $F_{D} = 0.075$,
for the filamentary flow shown
in Fig.~\ref{fig:9},
there is a narrow band noise signal.
(b) At $F_{D} = 0.1375$, a filamentary flow phase at a higher drive
also shows narrow band noise.
(c) At $F_{D}= 0.25$, in the fluctuating regime,
for low frequencies we find
$S(f) \propto 1/f^{\alpha}$ with $\alpha = 0.75$.
(d) At $F_{D} = 0.35$, the low frequency noise
power is reduced.}
\label{fig:11}
\end{figure}

The different phases produce signatures in the velocity
noise power spectra $S(f)$, obtained
from the time series of the conduction fluctuations shown in Fig.~\ref{fig:9}
according to
\begin{equation}
        S(f) = \frac{1}{2\pi}\left|\int V(t)e^{(-i2\pi ft)}dt\right|^2 \ .
\end{equation}
In computing the power spectrum, we discard the transient portion of the
velocity time series and only consider the time period during which
the system
has reached a steady state.
In Fig.~\ref{fig:11}(a) we plot $S(f)$
for the filamentary flow phase at
$F_{D} = 0.075$ from Fig.~\ref{fig:9}.
Here we find a narrow band signal with
multiple peaks produced by the periodicity of the flow of charges
through the q1D channels.
At $F_D=0.1375$ in Fig.~\ref{fig:11}(b),
a filamentary flow state at higher drives has narrow band peaks that
are shifted to higher
frequency since the charges are moving more rapidly through
the flow channels.
The power spectrum becomes
more complicated as additional channels of flow open, 
giving multiple different frequencies of flow.
In Fig.~\ref{fig:11}(c) we plot $S(f)$ in the
fluctuating liquid regime at $F_{D} = 0.25$.
The peaks associated with the periodic filamentary flow channels
are lost and the noise power at low frequencies assumes a
$1/f^\alpha$ form,
where the solid line is a fit with $\alpha = 0.75$.
This is close to the exponent values obtained
in recent noise measures on
sliding Wigner crystals, where
$\alpha = 0.6$ \cite{Brussarski18}.
As the drive increases further,
the low frequency noise power is reduced and the
spectrum at low frequencies becomes white,
as shown in Fig.~\ref{fig:11}(d)
for $F_{D} = 0.35$. When the moving smectic state forms
for even higher drives,
new narrow band noise peaks emerge that are associated
with the washboard frequency.
Washboard noise signals
for moving crystals or moving smectics
at higher drives have been studied in both simulation
and experiment for sliding charge density waves \cite{Gruner88},
superconducting vortices \cite{Reichhardt98,Togawa00,Kolton01,Okuma09},
and magnetic skyrmions \cite{Diaz17,Sato19,Sato20},
and have been predicted to occur for Wigner crystals \cite{Zhu94}.
Simulations of Wigner crystals moving over long
range disorder also produced a washboard signal \cite{Reichhardt01};
however, in the previous work
the narrow band noise in the filamentary
flow regime was not studied \cite{Reichhardt01}.

\begin{figure}
\includegraphics[width=\columnwidth]{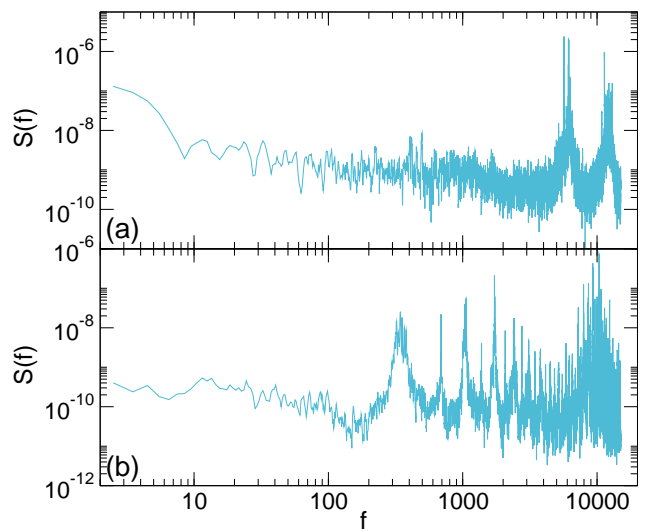}
\caption{$S(f)$
for the system from Figs.~\ref{fig:9} and \ref{fig:10} with $F=1.0$
and $F_p=0.5$.
(a) $F_{D} = 0.55$ in the moving smectic phase,
showing a narrow band signal at higher frequencies.
(b) $F_{D} = 1.0$, where there is
a pronounced
low frequency washboard signal.}
\label{fig:12}
\end{figure}

In Fig.~\ref{fig:12}(a) we plot
$S(f)$ for the system in
Fig. \ref{fig:10}
at $F_{D}  = 0.55$.
A narrow band noise signal
has emerged since the system is transitioning into
a moving smectic with
a small number of topological defects.
At $F_D=1.0$ in Fig.~\ref{fig:12}(b),
the ordering of the charge lattice has increased
and there is a stronger narrow band noise signal along with a
washboard signal at lower frequencies.

\begin{figure}
\includegraphics[width=\columnwidth]{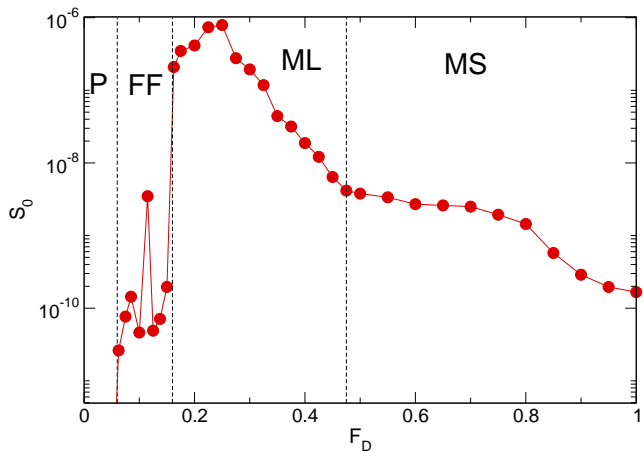}
\caption{
The noise power $S_{0}$ at a low fixed frequency of $f_{0} = 20$
vs $F_D$ for the system from Figs.~\ref{fig:9} and \ref{fig:10} with
$F=1.0$ and $F_p=0.5$.
$S_0$ is zero in the pinned (P) phase, low in the filamentary
flow (FF) phase, has a peak in the disordered moving liquid (ML) phase,
and drops again in the ordered moving smectic (MS) phase.}
\label{fig:13}
\end{figure}

In Fig.~\ref{fig:13} we plot
$S_{0}$, the noise power at a low fixed frequency of $f_{0} = 20$,
versus $F_D$ for the $F=1.0$ and $F_p=0.5$ system.
The noise power is 
low in the filamentary
flow phase, high in the fluctuating
moving liquid, and low in the dynamically reordered
moving smectic phase.
There is a peak in $S_0$ near $F_{D} = 0.25$. This falls
within the range $0.175 \leq F_D < 0.3$
in which the noise develops $1/f^{\alpha}$ characteristics, as shown in
Fig.~\ref{fig:11}(c).
The noise is white at higher drives of
$0.3 \leq F_{D} < 0.5$, similar to what is shown in Fig.~\ref{fig:11}(d).
For $0.5 < F_{D} < 0.8$,
a narrow band noise signal similar to that in Fig.~\ref{fig:12}(a)
appears, while for
$F_{D} \geq 0.8$, even fewer 
topological defects are present in the lattice and the
noise peaks sharpen, as shown in Fig.~\ref{fig:12}(b).
The peak in the noise power
illustrated in Fig.~\ref{fig:13}
is similar to low temperature
experimental results \cite{Brussarski18}
in which the noise power is small at low drives, reaches a strong
peak near depinning, and drops again at higher drives.

The appearance of low frequency
narrow band noise near depinning and broad band noise
at higher drives has also been observed for
Wigner crystals and electron liquid crystals \cite{Cooper03,Sun22}.
In the work of Sun {\it et al.} \cite{Sun22},
transitions among ordered and disordered
phases occur as a function of
drive, and the ordered phases exhibit narrow band noise features.
This could indicate that there are regimes 
of stable filament flow
interspersed with fluctuating flow regimes,
and that at high drives the system
dynamically reorders.
In our case, in the filamentary flow regime the system generally jumps from one
ordered state to another,
but there are situations in which
the charges remain in a fluctuating
state between the stable filamentary regimes.
This occurs more frequently for continuous driving, as shown in
Fig.~\ref{fig:4}(c).
If temperature is added,
additional fluctuating flow phases might emerge.
The narrow band noise in the filamentary phase does not arise from
the presence of strong ordering, for the filamentary phase is
structurally disordered. Instead, since the q1D filaments are stable,
motion along the filaments repeats very reliably as a function of time.
One question is whether
the fluctuating regimes
are transient, meaning that the system might order after a sufficiently
long time interval, or whether they are in fact stable, meaning that
it is only possible for the system to reach an ordered state for
higher drives.
In systems where charged stripes or bubbles
can arise,
such as in reentrant quantum Hall systems,
experimental measurements show
that there is narrow band
noise \cite{Bennaceur18}.
In experiments on 2D electron systems,
nonlinear current-voltage
curves
interpreted as signatures of
the depinning of electron nematics or smectics are also associated
with $1/f$ noise
\cite{Qian17}.

\section{Varied Disorder Strength}

\begin{figure}
\includegraphics[width=\columnwidth]{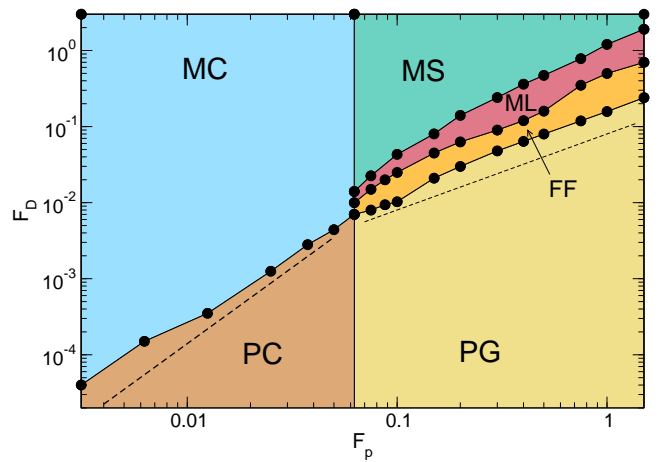}
\caption{
Dynamic phase diagram as a function of $F_{D}$ vs
pinning strength $F_{p}$ for a system with $F = 1.0$.
At low $F_{p}$, there is a pinned crystal
(PC, light brown) that depins elastically 
into a moving crystal (MC, light blue).
For higher $F_{p}$,
the system is initially in a pinned glass
state (PG, yellow) that depins plastically into the filamentary
flow regime (FF, orange). At higher drives, there is a transition to
the disordered moving liquid (ML, pink) state and a reordering into
the moving smectic (MS, green) state.
Fits to the depinning curve are shown as dashed lines:
left, $F_{c} \propto F^{2}_{p}$; right, 
$F_{c} \propto F_{p}$.}
\label{fig:14}
\end{figure}

We next consider the effects of changing the pinning strength $F_p$.
We first focus on a system with
fixed $F = 1.0$ and measure the
velocity-force curves
and $P_{6}$ versus $F_D$.
From this data, we construct
the dynamic phase diagram as a function of
$F_{D}$ versus $F_{p}$
plotted in Fig.~\ref{fig:14}.
For $F_{p} \geq 0.075$,
the ground state at $F_{D} = 0.0$ is a disordered pinned glass
since the pinning is strong enough to induce
the formation of topological defects in the
Wigner crystal.
This pinned glass
depins into the filamentary flow phase,
which exhibits narrow band noise.
At higher drives, there is a transition
into the fluctuating moving liquid phase with
$1/f$ noise, followed by dynamical reordering into a moving smectic.
The depinning threshold for the pinned glass increases
linearly with $F_p$, as indicated by the
dashed line showing a fit to $F_c \propto F_p$, similar to what
has been observed in previous work on plastic depinning transitions
\cite{Reichhardt17}.
The drives at which the system transitions
from the filamentary flow phase to the disordered moving liquid
and from the moving liquid to the moving smectic also increase
linearly with increasing $F_p$.
For $F_{p} < 0.075$, the disorder is weak enough that no topological defects
form in the pinned state, and
a pinned crystal appears
that depins elastically to a moving crystal.
In this regime, the filamentary flow and disordered moving liquid phases
are absent.
In general, when the pinning is strong,
some topological defects form in the moving state, and since these defects
align so as to glide in the driving direction, the system is best described
as a moving smectic. In contrast,
when the disorder is weak, there are no topological defects
and the system is described as a moving crystal.
The depinning threshold in the elastic  depinning regime obeys
$F_{c} \propto F^{2}_{p}$,
as expected for collective or elastic depinning \cite{Reichhardt17}.

\begin{figure}
\includegraphics[width=\columnwidth]{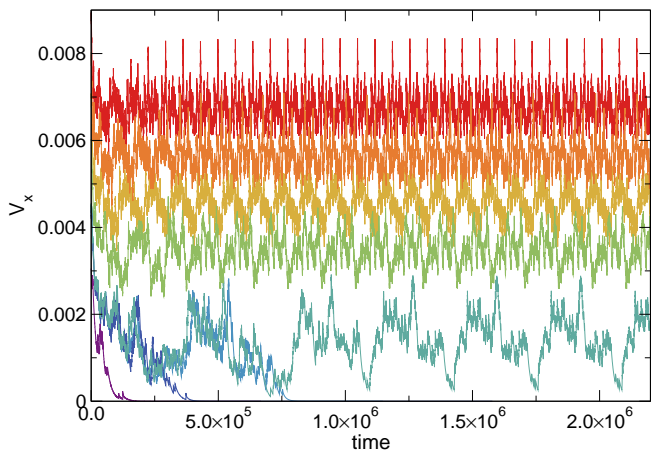}
\caption{
Time series of the velocity $V_x$ in the elastic depinning regime
for the system in Fig.~\ref{fig:14} with $F=1.0$ at $F_p=0.05$ and  
$F_{D} = 0.0035$ (violet),
0.00425 (dark blue),
0.00435 (light blue),
0.0044  (teal),
0.006 (green),
0.007 (yellow),
0.008 (orange), and  $0.009$ (red).
For $F_D < 0.044 $ the system evolves to a pinned state.
For $F_{D} \geq 0.044$,
$V_x$ develops
a washboard or periodic noise signal.
}
\label{fig:15}
\end{figure}

In the elastic depinning regime,
there is no filamentary flow and the depinning occurs
in a single step, giving transport and noise signatures that are
very distinct from those found for the plastic depinning
transition where filamentary flow occurs.
In Fig.~\ref{fig:15} we plot the
velocity time series $V_x$
for the system in
Fig.~\ref{fig:14}
at $F_p=0.05$ over a range of $F_D$ from $F_D=0.0035$
to $F_D=0.009$ that spans the depinning threshold.
For $F_{p} < 0.044$, the flow is transient and the system
settles into a pinned state.
For $F_{D} \geq 0.044$, the motion persists and
develops a complex periodic signal.
As $F_D$ increases, the magnitude of $V_x$ and the frequency of its
oscillations both increase.
This result indicates that in the elastic depinning regime,
a washboard signal emerges above the depinning threshold.

\begin{figure}
\includegraphics[width=\columnwidth]{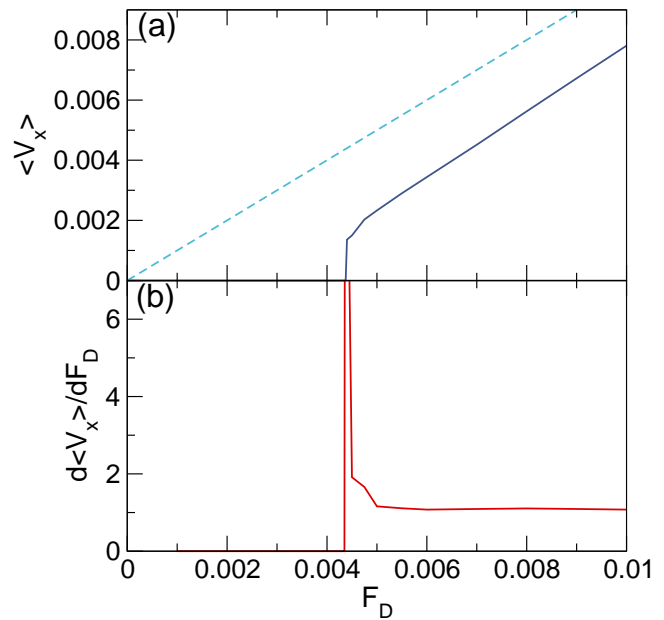}
\caption{
The average velocity $\langle V_x\rangle$ (solid line) vs $F_D$
for $F_p=0.05$ in the elastic depinning regime
of the system from Fig.~\ref{fig:14} and
Fig.~\ref{fig:15} with $F=1.0$. The dashed
line is the pin free result.
(b) The corresponding $d\langle V_x\rangle/dF_{D}$ versus $F_D$ has
a single sharp peak at depinning.}
\label{fig:16}
\end{figure}

\begin{figure}
\includegraphics[width=\columnwidth]{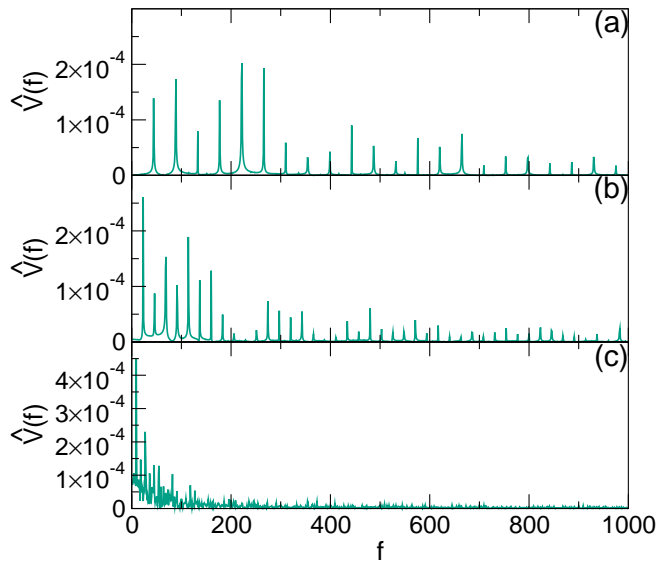}
\caption{
The Fourier transform ${\hat V}(f)$
of the velocity time series $V_x$ for the system in
Fig.~\ref{fig:15} with $F=1.0$ and $F_p=0.05$
at $F_{D} =$
(a) 0.9,
(b) 0.7, and
(c) $0.044$.
The narrow band noise peaks shift to higher frequencies with
increasing $F_{D}$.}
\label{fig:17}
\end{figure}

Figure~\ref{fig:16}(a) shows $\langle V_x\rangle$
versus $F_{D}$ for the
system in Fig.~\ref{fig:15},
where the dashed line is the pin free response.
The corresponding $d\langle V_x\rangle/dF_D$ versus $F_D$ in
Fig.~\ref{fig:16}(b)
has a single sharp peak at depinning, in contrast to the double
peak that appears in a two step plastic depinning process.
When the depinning is elastic,
both the filamentary flow phase and switching events are absent.
In Fig.~\ref{fig:17}(a,b,c) we plot the Fourier transform ${\hat V}(f)$ of the
velocity time series $V_x$ for the system in Fig.~\ref{fig:15} at
$F_{D} = 0.9$, 0.7, and $0.044$, respectively.
The narrow band noise peaks shift to higher frequencies as $F_D$ increases.

\begin{figure}
\includegraphics[width=\columnwidth]{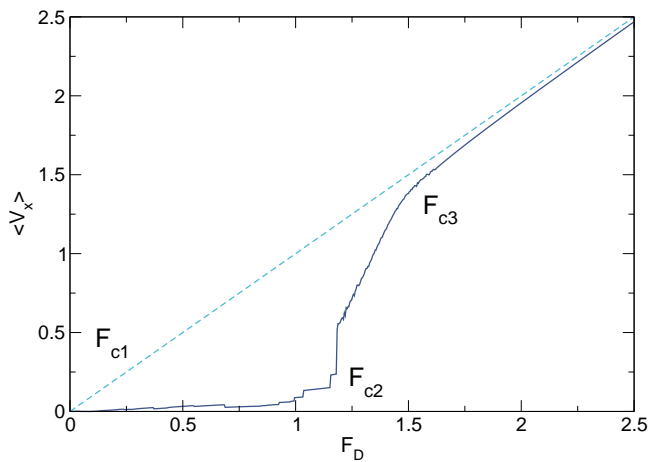}
\caption{
$\langle V_x\rangle$
vs $F_{D}$ for a system with $F_{p} = 1.5$ and $F = 0.4$.
The drive is increased in increments of
$\Delta F_{D} = 0.004$ and we
spend 888400 simulation time steps at each increment to obtain
the average velocity value.
$F_{c1}$ is the transition from pinned to filamentary flow,
$F_{c2}$ denotes the transition from filamentary to
disordered fluctuating flow,
and $F_{c3}$ is the transition from disordered to linear flow.}
\label{fig:18}
\end{figure}

\begin{figure}
\includegraphics[width=\columnwidth]{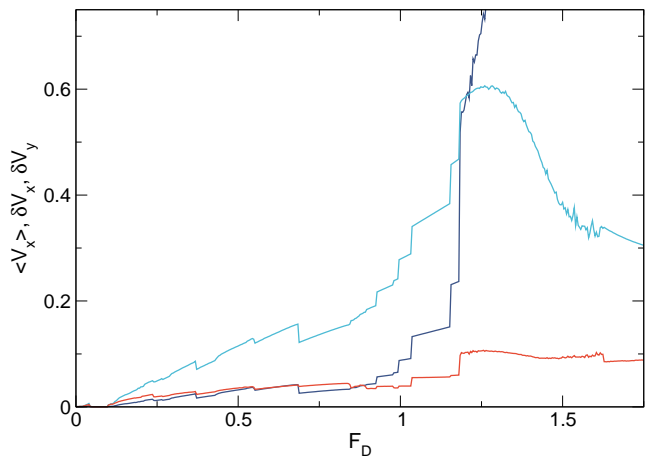}
\caption{
$\langle V_x\rangle$ (dark blue),
$\delta V_{x}$ (light blue), and $\delta V_{y}$ (red) vs $F_D$
from the filamentary regime through
the transition to the plastic flow regime for the system in
Fig.~\ref{fig:18} with $F_p=1.5$ and $F=0.4$.}
\label{fig:19}
\end{figure}

For larger $F_{p}$, the velocity-force curves
exhibit stronger cusps at the transitions
among the different phases.
Recent experimental
work identified a regime in which there is
a clear two step depinning process, with an
initial transition to a nonlinear regime followed by a second transition
to a phase in which the velocity increases linearly with increasing drive
\cite{Brussarski18}.
In Fig.~\ref{fig:18} we plot $\langle V_x\rangle$ versus
$F_{D}$ for a system with $F_{p} = 1.5$ and $F = 0.4$. To obtain
this curve,
we increase the driving force in increments of
$\Delta F_{D} = 0.004$ and spend 
888400 simulation time steps on each increment in order to obtain
an average velocity and produce a smoother velocity-force curve.
As Fig.~\ref{fig:18} shows, there is an extended regime of filamentary flow.
The transition from the pinned state to the filamentary flow state
is labeled $F_{c1}$, the transition from filamentary flow to
the disordered fluctuating flow state is marked $F_{c2}$, and
the transition to the linear flow state is denoted $F_{c3}$.
In the experimental work of Ref.~\cite{Brussarski18},
only $F_{c2}$ and $F_{c3}$ were observed. It may be possible  that
there is a filamentary flow phase that is
not resolvable experimentally
due to the large fluctuations in the data.
In Fig.~\ref{fig:19} we plot $\langle V_x\rangle$,
$\delta V_{x}$, and $\delta V_{y}$ versus $F_D$
from the filamentary regime through the transition to plastic flow regime.
Within the filamentary regime,
a series of upward and downward jumps occur
in $\langle V_x\rangle$
corresponding to the opening and closing of
individual flow channels.
These jumps are also
correlated with jumps in $\delta V_{x}$ and $\delta V_{y}$.
In general,
$\delta V_{x}$ is larger than $\delta V_{y}$,
and a large peak in $\delta V_{x}$ appears at the transition to the plastic
flow regime.

\begin{figure}
\includegraphics[width=\columnwidth]{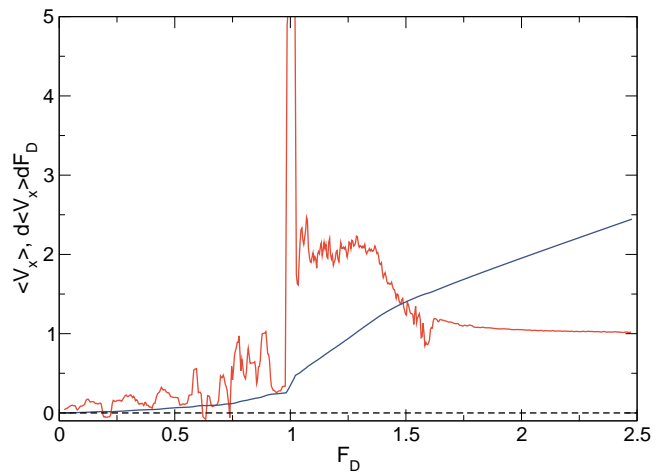}
\caption{$\langle V_x\rangle$ (blue) and $d\langle V_x\rangle/dF_{D}$ (red)
vs $F_{D}$ for a system with $F_{p} = 1.5$ and $F = 0.8$.
The dashed line indicates that $d\langle V_x\rangle/dF_D$ makes excursions
below zero.}
\label{fig:20}
\end{figure}

As we vary the filling at $F_p=1.5$,
we find similar velocity-force and velocity fluctuation curves.
For example,
in Fig.~\ref{fig:20} we plot $\langle V_x\rangle$ and
$d\langle V_x\rangle/dF_{D}$ versus $F_{D}$ for
a system with $F = 0.8$.
The filamentary flow phase is visible as
a region of positive and negative peaks in
$d\langle V_x\rangle/dF_{D}$.
The transition to the nonlinear
flow regime is accompanied by a
large peak in $d\langle V_x\rangle/dF_D$,
followed by small dip
after which $d\langle V_x\rangle/dF_{D}$ approaches $1.0$.
These results suggest that the two step depinning
process observed in Ref.~\cite{Brussarski18} is likely associated with
a stronger pinning regime.

\section{Avalanches}

\begin{figure}
\includegraphics[width=\columnwidth]{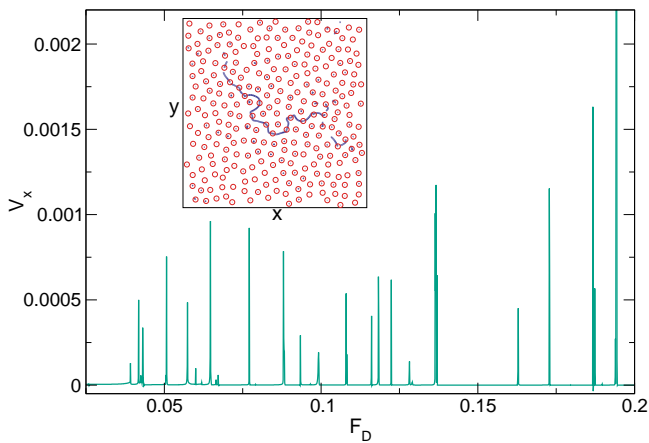}
\caption{
$V_x$ averaged over a very short time interval
vs $F_{D}$ for a system with $F = 0.4$ and $F_{p}= 0.5$ in the pinned
regime showing a series of avalanches. Inset: Image of the
electron locations (red circles) and trajectories (lines)
showing an example of an avalanche in which the motion occurs along
a q1D chain.
}
\label{fig:21}
\end{figure}

\begin{figure}
\includegraphics[width=\columnwidth]{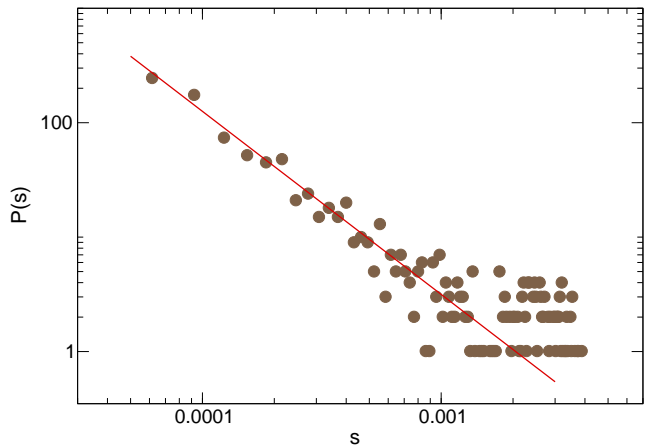}
\caption{
Distribution $P(s)$ of the avalanche sizes measured in terms of the
magnitude of $V_x$ during the avalanche event for 100 disorder
realizations of the system from Fig.~\ref{fig:20} with $F=0.4$ and
$F_p=0.5$ in the pinned regime.
The dashed line is a power law fit
to $P(s) \propto s^{-\tau}$
with $\tau = 1.6$.}
\label{fig:22}
\end{figure}

As noted in Fig.~\ref{fig:1}, there are clear jumps in
$\delta V_x$ and $\delta V_y$ for increasing $F_D$
within the pinned regime
that are associated with charge avalanches.
In general, the avalanches are the most prominent in
parameter windows where filamentary flow can occur.
In Fig.~\ref{fig:21} we plot
$V_x$ averaged over a very short time interval
versus $F_{D}$ for a system with $F = 0.4$ and $F_{p}= 0.5$.
The inset shows an
image of one of the avalanches in which
the motion occurs along a q1D
chain.
This motion resembles what is found in the filamentary flow phase, but the
duration of the motion is finite, indicating that
the avalanches can be viewed as q1D excitations.
To obtain better statistics on the
avalanche behavior, we performed
a series of 100 different disorder realizations
with different pinning site locations and
swept the value of $F_{D}$ as in Fig.~\ref{fig:21}.
We define the size $s$ of the avalanche to be the magnitude of $V_x$.
In Fig.~\ref{fig:22} we plot the avalanche size distribution
$P(s)$ on a log-log scale.
The behavior is consistent with $P(s) \propto s^{-\tau}$,
where the solid line is a fit with $\tau = 1.6$.
Previous work on avalanches
in three dimensional Coulomb systems obtained an
avalanche exponent of $\tau = 1.5$ \cite{Palassini12}.
Among
systems
of particles moving over random pinning, the one
with q1D avalanches that is the 
closest to what we observe here
is vortices in type II superconductors, where
avalanches in the strong pinning regime
form q1D like chains and the avalanche sizes are power law
distributed with
exponents ranging from $\tau=2.0$ to $\tau=1.6$
\cite{Olson97,Bassler98}.
In a magnetic skyrmion system, avalanches were observed
with an exponent of $\tau = 1.5$ \cite{Diaz18}.
These results indicate that the behavior of charges moving over random
disorder is similar to that of
superconducting vortices and
magnetic skyrmions. This is reasonable since all of these systems
consist of
an assembly of driven particles
moving over quenched disorder. In the case of
the Wigner
crystal, the interactions are Coulomb
in form, while for superconducting vortices and
magnetic skyrmions, the interactions are
closer to screened Coulomb.

For lower values of $F$, the avalanches do not
occur along q1D chains but rather
consist of single hops,
and the avalanche distributions show a characteristic  peak associated with
the velocity of a single electron
jumping from one pinning site to the next.
Avalanches also occur below depinning in the elastic depinning regime,
but these avalanches are 2D in nature
and the statistics are much more difficult to obtain.
Avalanches in charge ordering systems
have been studied in the presence of strong pinning, and
the largest number of
avalanches occur near what are believed
to be critical points as a function of temperature
\cite{Baity18}.
Future directions include testing whether
the avalanche shapes show scaling, as
found in crackling  noise
systems \cite{Diaz18,Perkovic95}.

\section{Discussion}

In the experimental work
of Ref.~\cite{Brussarski18}, the current-voltage curves
become much more rounded
with increasing temperature.
This can be interpreted as resulting
from an increase in
thermal creep.
Increasing the temperature
also destroyed the peak
in the noise power at depinning
and led to an overall reduction in the noise power.
Future work could examine such thermal effects in
greater detail.
For example, in the filamentary flow regime,
the transport could develop additional thermal switching
effects in which thermally induced random hops could
cause some channels to open or close
even at a fixed drive, giving rise to
telegraph noise.
It would also be interesting to
determine whether
the strong
narrow band noise signals in the filamentary or moving crystal phases are
robust against thermal fluctuations.

In our system,
the filamentary flow channels are one particle wide and
are essentially
one dimensional.
Other flow morphologies are possible,
such as large scale mesoscopic rivers
in which there is a mixture of fluid
and solid states, so that large rivers wind
though the sample between pinned
islands.
This type of flow could arise if large scale
heterogeneities are present, causing
the disorder strength to vary over
length scales that are multiple
lattice constants of the periodicity of the Wigner crystal.
Much
larger scale simulations would be needed to explore such regimes, which
could include tens to hundreds of thousands of charges.

Another question is the role of screening,
which could modify the electron-electron interactions.
If screening is important,
the
interactions would be better described as Yukawa
or screened Coulomb in form, $\exp(-\kappa r)/r$.
Interactions of this type
could produce behavior more similar
to that of colloidal assemblies moving over random
disorder, where plastic to elastic depinning transitions
have been studied \cite{Reichhardt02,Pertsinidis08}.
Other directions would be to explore critical behaviors
in the velocity scaling or transient effects near the
Wigner crystal depinning transition,
and compare these to
the scaling found in other systems, such as
superconducting vortices
\cite{Reichhardt09,Kaji22}.

\section{Summary} 

We have investigated the nonlinear dynamics of Wigner crystals driven
over random disorder
by measuring the velocity-force curves,
the standard deviation of the velocities,
and the fraction of six-fold coordinated charges. For strong
disorder, the system
forms a pinned Wigner glass that can depin 
into a filamentary flow state where
the motion occurs in well defined quasi-one-dimensional channels.
As the drive increases, these channels
can open or close, leading to switching events that have either positive
or negative differential conductivity.
In regimes where there is stable filamentary flow, the
noise has a narrow band character, and at the transition to
fluctuating filaments or disordered liquid flow, the noise
becomes broad band and the noise power at low frequencies grows
large.
At higher drives, the
system enters a continuously fluctuating state
in which pinned and mobile electrons coexist and there is a
rapid exchange between the two.
In this case there is strong broad brand noise of
$1/f^{\alpha}$ form with $\alpha = 0.7$.
At even higher drives, the system can reorder into a moving
smectic where most of the charges have six neighbors
and the noise shows a washboard signal.
This indicates that for increasing drive,
there is a regime of narrow band noise for the initial flow,
followed by $1/f$ noise and a peak of the noise power in the
intermediate plastic flow regime,
and then a reappearance
of narrow band noise at high drives.
For weak quenched disorder, the system
forms a pinned crystal that undergoes
a single depinning transition into a moving crystal
with a strong narrow band noise signal.
Within the strong pinning regime, there can
be a two or three step depinning process,
where there is enhanced filamentary flow
with regions of negative differential
conductivity, 
a disordered flow regime, and finally a transition to a
regime in which 
the velocity increases linearly with drive.
These phases produce clear signatures in the
differential conductivity curves.
Many of the transport features we observe are in agreement
with recent transport studies
on Wigner crystals in a
regime where stripes or bubble like charge ordered phases could be
occurring.
Finally, we find that
below the conduction threshold,
as the drive is increased
there can be large scale rearrangements within
the pinned phase in the form of avalanches.
These avalanches have quasi-one-dimensional characteristics and their
sizes are power law distributed with an exponent of
$\tau = 1.6$, similar to what is found for avalanches in driven
vortices in type-II superconductors.

\begin{acknowledgments}
We gratefully acknowledge the support of the U.S. Department of
Energy through the LANL/LDRD program for this work.
This work was supported by the US Department of Energy through
the Los Alamos National Laboratory.  Los Alamos National Laboratory is
operated by Triad National Security, LLC, for the National Nuclear Security
Administration of the U. S. Department of Energy (Contract No. 892333218NCA000001).

\end{acknowledgments}

\bibliography{mybib}

\begin{thebibliography}{89}
\expandafter\ifx\csname natexlab\endcsname\relax\def\natexlab#1{#1}\fi
\expandafter\ifx\csname bibnamefont\endcsname\relax
  \def\bibnamefont#1{#1}\fi
\expandafter\ifx\csname bibfnamefont\endcsname\relax
  \def\bibfnamefont#1{#1}\fi
\expandafter\ifx\csname citenamefont\endcsname\relax
  \def\citenamefont#1{#1}\fi
\expandafter\ifx\csname url\endcsname\relax
  \def\url#1{\texttt{#1}}\fi
\expandafter\ifx\csname urlprefix\endcsname\relax\def\urlprefix{URL }\fi
\providecommand{\bibinfo}[2]{#2}
\providecommand{\eprint}[2][]{\url{#2}}

\bibitem[{\citenamefont{Reichhardt and Reichhardt}(2017)}]{Reichhardt17}
\bibinfo{author}{\bibfnamefont{C.}~\bibnamefont{Reichhardt}} \bibnamefont{and}
  \bibinfo{author}{\bibfnamefont{C.~J.~O.} \bibnamefont{Reichhardt}},
  \bibinfo{journal}{Rep. Prog. Phys.} \textbf{\bibinfo{volume}{80}},
  \bibinfo{pages}{026501} (\bibinfo{year}{2017}).

\bibitem[{\citenamefont{Bhattacharya and Higgins}(1993)}]{Bhattacharya93}
\bibinfo{author}{\bibfnamefont{S.}~\bibnamefont{Bhattacharya}}
  \bibnamefont{and} \bibinfo{author}{\bibfnamefont{M.~J.}
  \bibnamefont{Higgins}}, \bibinfo{journal}{Phys. Rev. Lett.}
  \textbf{\bibinfo{volume}{70}}, \bibinfo{pages}{2617} (\bibinfo{year}{1993}).

\bibitem[{\citenamefont{Reichhardt and Olson}(2002)}]{Reichhardt02}
\bibinfo{author}{\bibfnamefont{C.}~\bibnamefont{Reichhardt}} \bibnamefont{and}
  \bibinfo{author}{\bibfnamefont{C.~J.} \bibnamefont{Olson}},
  \bibinfo{journal}{Phys. Rev. Lett.} \textbf{\bibinfo{volume}{89}},
  \bibinfo{pages}{078301} (\bibinfo{year}{2002}).

\bibitem[{\citenamefont{Pertsinidis and Ling}(2008)}]{Pertsinidis08}
\bibinfo{author}{\bibfnamefont{A.}~\bibnamefont{Pertsinidis}} \bibnamefont{and}
  \bibinfo{author}{\bibfnamefont{X.~S.} \bibnamefont{Ling}},
  \bibinfo{journal}{Phys. Rev. Lett.} \textbf{\bibinfo{volume}{100}},
  \bibinfo{pages}{028303} (\bibinfo{year}{2008}).

\bibitem[{\citenamefont{Vanossi et~al.}(2013)\citenamefont{Vanossi, Manini,
  Urbakh, Zapperi, and Tosatti}}]{Vanossi13}
\bibinfo{author}{\bibfnamefont{A.}~\bibnamefont{Vanossi}},
  \bibinfo{author}{\bibfnamefont{N.}~\bibnamefont{Manini}},
  \bibinfo{author}{\bibfnamefont{M.}~\bibnamefont{Urbakh}},
  \bibinfo{author}{\bibfnamefont{S.}~\bibnamefont{Zapperi}}, \bibnamefont{and}
  \bibinfo{author}{\bibfnamefont{E.}~\bibnamefont{Tosatti}},
  \bibinfo{journal}{Rev. Mod. Phys.} \textbf{\bibinfo{volume}{85}},
  \bibinfo{pages}{529} (\bibinfo{year}{2013}).

\bibitem[{\citenamefont{Reichhardt et~al.}(2015)\citenamefont{Reichhardt, Ray,
  and Reichhardt}}]{Reichhardt15}
\bibinfo{author}{\bibfnamefont{C.}~\bibnamefont{Reichhardt}},
  \bibinfo{author}{\bibfnamefont{D.}~\bibnamefont{Ray}}, \bibnamefont{and}
  \bibinfo{author}{\bibfnamefont{C.~J.~O.} \bibnamefont{Reichhardt}},
  \bibinfo{journal}{Phys. Rev. Lett.} \textbf{\bibinfo{volume}{114}},
  \bibinfo{pages}{217202} (\bibinfo{year}{2015}).

\bibitem[{\citenamefont{Jiang et~al.}(2017)\citenamefont{Jiang, Zhang, Yu,
  Zhang, Wang, Jungfleisch, Pearson, Cheng, Heinonen, Wang et~al.}}]{Jiang17}
\bibinfo{author}{\bibfnamefont{W.}~\bibnamefont{Jiang}},
  \bibinfo{author}{\bibfnamefont{X.}~\bibnamefont{Zhang}},
  \bibinfo{author}{\bibfnamefont{G.}~\bibnamefont{Yu}},
  \bibinfo{author}{\bibfnamefont{W.}~\bibnamefont{Zhang}},
  \bibinfo{author}{\bibfnamefont{X.}~\bibnamefont{Wang}},
  \bibinfo{author}{\bibfnamefont{M.~B.} \bibnamefont{Jungfleisch}},
  \bibinfo{author}{\bibfnamefont{J.~E.} \bibnamefont{Pearson}},
  \bibinfo{author}{\bibfnamefont{X.}~\bibnamefont{Cheng}},
  \bibinfo{author}{\bibfnamefont{O.}~\bibnamefont{Heinonen}},
  \bibinfo{author}{\bibfnamefont{K.~L.} \bibnamefont{Wang}},
  \bibnamefont{et~al.}, \bibinfo{journal}{Nature Phys.}
  \textbf{\bibinfo{volume}{13}}, \bibinfo{pages}{162} (\bibinfo{year}{2017}).

\bibitem[{\citenamefont{S\'andor et~al.}(2017)\citenamefont{S\'andor, Lib\'al,
  Reichhardt, and Olson~Reichhardt}}]{Sandor17a}
\bibinfo{author}{\bibfnamefont{C.}~\bibnamefont{S\'andor}},
  \bibinfo{author}{\bibfnamefont{A.}~\bibnamefont{Lib\'al}},
  \bibinfo{author}{\bibfnamefont{C.}~\bibnamefont{Reichhardt}},
  \bibnamefont{and} \bibinfo{author}{\bibfnamefont{C.~J.}
  \bibnamefont{Olson~Reichhardt}}, \bibinfo{journal}{Phys. Rev. E}
  \textbf{\bibinfo{volume}{95}}, \bibinfo{pages}{032606}
  (\bibinfo{year}{2017}).

\bibitem[{\citenamefont{Morin et~al.}(2017)\citenamefont{Morin, Desreumaux,
  Caussin, and Bartolo}}]{Morin17}
\bibinfo{author}{\bibfnamefont{A.}~\bibnamefont{Morin}},
  \bibinfo{author}{\bibfnamefont{N.}~\bibnamefont{Desreumaux}},
  \bibinfo{author}{\bibfnamefont{J.-B.} \bibnamefont{Caussin}},
  \bibnamefont{and} \bibinfo{author}{\bibfnamefont{D.}~\bibnamefont{Bartolo}},
  \bibinfo{journal}{Nature Phys.} \textbf{\bibinfo{volume}{13}},
  \bibinfo{pages}{63} (\bibinfo{year}{2017}).

\bibitem[{\citenamefont{Giamarchi and Le~Doussal}(1996)}]{Giamarchi96}
\bibinfo{author}{\bibfnamefont{T.}~\bibnamefont{Giamarchi}} \bibnamefont{and}
  \bibinfo{author}{\bibfnamefont{P.}~\bibnamefont{Le~Doussal}},
  \bibinfo{journal}{Phys. Rev. Lett.} \textbf{\bibinfo{volume}{76}},
  \bibinfo{pages}{3408} (\bibinfo{year}{1996}).

\bibitem[{\citenamefont{Balents et~al.}(1998)\citenamefont{Balents, Marchetti,
  and Radzihovsky}}]{Balents98}
\bibinfo{author}{\bibfnamefont{L.}~\bibnamefont{Balents}},
  \bibinfo{author}{\bibfnamefont{M.~C.} \bibnamefont{Marchetti}},
  \bibnamefont{and}
  \bibinfo{author}{\bibfnamefont{L.}~\bibnamefont{Radzihovsky}},
  \bibinfo{journal}{Phys. Rev. B} \textbf{\bibinfo{volume}{57}},
  \bibinfo{pages}{7705} (\bibinfo{year}{1998}).

\bibitem[{\citenamefont{Olson et~al.}(1998)\citenamefont{Olson, Reichhardt, and
  Nori}}]{Olson98a}
\bibinfo{author}{\bibfnamefont{C.~J.} \bibnamefont{Olson}},
  \bibinfo{author}{\bibfnamefont{C.}~\bibnamefont{Reichhardt}},
  \bibnamefont{and} \bibinfo{author}{\bibfnamefont{F.}~\bibnamefont{Nori}},
  \bibinfo{journal}{Phys. Rev. Lett.} \textbf{\bibinfo{volume}{81}},
  \bibinfo{pages}{3757} (\bibinfo{year}{1998}).

\bibitem[{\citenamefont{Fisher}(1998)}]{Fisher98}
\bibinfo{author}{\bibfnamefont{D.~S.} \bibnamefont{Fisher}},
  \bibinfo{journal}{Phys. Rep.} \textbf{\bibinfo{volume}{301}},
  \bibinfo{pages}{113} (\bibinfo{year}{1998}).

\bibitem[{\citenamefont{Fily et~al.}(2010)\citenamefont{Fily, Olive, Di~Scala,
  and Soret}}]{Fily10}
\bibinfo{author}{\bibfnamefont{Y.}~\bibnamefont{Fily}},
  \bibinfo{author}{\bibfnamefont{E.}~\bibnamefont{Olive}},
  \bibinfo{author}{\bibfnamefont{N.}~\bibnamefont{Di~Scala}}, \bibnamefont{and}
  \bibinfo{author}{\bibfnamefont{J.~C.} \bibnamefont{Soret}},
  \bibinfo{journal}{Phys. Rev. B} \textbf{\bibinfo{volume}{82}},
  \bibinfo{pages}{134519} (\bibinfo{year}{2010}).

\bibitem[{\citenamefont{Gr{\o}nbech-Jensen
  et~al.}(1996)\citenamefont{Gr{\o}nbech-Jensen, Bishop, and Dom{\'
  \i}nguez}}]{GronbechJensen96}
\bibinfo{author}{\bibfnamefont{N.}~\bibnamefont{Gr{\o}nbech-Jensen}},
  \bibinfo{author}{\bibfnamefont{A.~R.} \bibnamefont{Bishop}},
  \bibnamefont{and} \bibinfo{author}{\bibfnamefont{D.}~\bibnamefont{Dom{\'
  \i}nguez}}, \bibinfo{journal}{Phys. Rev. Lett.}
  \textbf{\bibinfo{volume}{76}}, \bibinfo{pages}{2985} (\bibinfo{year}{1996}).

\bibitem[{\citenamefont{Reichhardt et~al.}(1998)\citenamefont{Reichhardt,
  Olson, and Nori}}]{Reichhardt98}
\bibinfo{author}{\bibfnamefont{C.}~\bibnamefont{Reichhardt}},
  \bibinfo{author}{\bibfnamefont{C.~J.} \bibnamefont{Olson}}, \bibnamefont{and}
  \bibinfo{author}{\bibfnamefont{F.}~\bibnamefont{Nori}},
  \bibinfo{journal}{Phys. Rev. B} \textbf{\bibinfo{volume}{58}},
  \bibinfo{pages}{6534} (\bibinfo{year}{1998}).

\bibitem[{\citenamefont{Gr\"uner}(1988)}]{Gruner88}
\bibinfo{author}{\bibfnamefont{G.}~\bibnamefont{Gr\"uner}},
  \bibinfo{journal}{Rev. Mod. Phys.} \textbf{\bibinfo{volume}{60}},
  \bibinfo{pages}{1129} (\bibinfo{year}{1988}).

\bibitem[{\citenamefont{Besseling et~al.}(2005)\citenamefont{Besseling, Kes,
  Drose, and Vinokur}}]{Besseling05}
\bibinfo{author}{\bibfnamefont{R.}~\bibnamefont{Besseling}},
  \bibinfo{author}{\bibfnamefont{P.~H.} \bibnamefont{Kes}},
  \bibinfo{author}{\bibfnamefont{T.}~\bibnamefont{Drose}}, \bibnamefont{and}
  \bibinfo{author}{\bibfnamefont{V.~M.} \bibnamefont{Vinokur}},
  \bibinfo{journal}{New J. Phys.} \textbf{\bibinfo{volume}{7}},
  \bibinfo{pages}{71} (\bibinfo{year}{2005}).

\bibitem[{\citenamefont{Bag et~al.}(2017)\citenamefont{Bag, Shaw, Banerjee,
  Majumdar, Sood, and Grover}}]{Bag17}
\bibinfo{author}{\bibfnamefont{B.}~\bibnamefont{Bag}},
  \bibinfo{author}{\bibfnamefont{G.}~\bibnamefont{Shaw}},
  \bibinfo{author}{\bibfnamefont{S.~S.} \bibnamefont{Banerjee}},
  \bibinfo{author}{\bibfnamefont{S.}~\bibnamefont{Majumdar}},
  \bibinfo{author}{\bibfnamefont{A.~K.} \bibnamefont{Sood}}, \bibnamefont{and}
  \bibinfo{author}{\bibfnamefont{A.~K.} \bibnamefont{Grover}},
  \bibinfo{journal}{Sci. Rep.} \textbf{\bibinfo{volume}{7}},
  \bibinfo{pages}{5531} (\bibinfo{year}{2017}).

\bibitem[{\citenamefont{Cooper et~al.}(2003)\citenamefont{Cooper, Eisenstein,
  Pfeiffer, and West}}]{Cooper03}
\bibinfo{author}{\bibfnamefont{K.~B.} \bibnamefont{Cooper}},
  \bibinfo{author}{\bibfnamefont{J.~P.} \bibnamefont{Eisenstein}},
  \bibinfo{author}{\bibfnamefont{L.~N.} \bibnamefont{Pfeiffer}},
  \bibnamefont{and} \bibinfo{author}{\bibfnamefont{K.~W.} \bibnamefont{West}},
  \bibinfo{journal}{Phys. Rev. Lett.} \textbf{\bibinfo{volume}{90}},
  \bibinfo{pages}{226803} (\bibinfo{year}{2003}).

\bibitem[{\citenamefont{Qian et~al.}(2017)\citenamefont{Qian, Nakamura,
  Fallahi, Gardner, and Manfra}}]{Qian17}
\bibinfo{author}{\bibfnamefont{Q.}~\bibnamefont{Qian}},
  \bibinfo{author}{\bibfnamefont{J.}~\bibnamefont{Nakamura}},
  \bibinfo{author}{\bibfnamefont{S.}~\bibnamefont{Fallahi}},
  \bibinfo{author}{\bibfnamefont{G.~C.} \bibnamefont{Gardner}},
  \bibnamefont{and} \bibinfo{author}{\bibfnamefont{M.~J.}
  \bibnamefont{Manfra}}, \bibinfo{journal}{Nature Commun.}
  \textbf{\bibinfo{volume}{8}}, \bibinfo{pages}{1536} (\bibinfo{year}{2017}).

\bibitem[{\citenamefont{Marley et~al.}(1995)\citenamefont{Marley, Higgins, and
  Bhattacharya}}]{Marley95}
\bibinfo{author}{\bibfnamefont{A.~C.} \bibnamefont{Marley}},
  \bibinfo{author}{\bibfnamefont{M.~J.} \bibnamefont{Higgins}},
  \bibnamefont{and}
  \bibinfo{author}{\bibfnamefont{S.}~\bibnamefont{Bhattacharya}},
  \bibinfo{journal}{Phys. Rev. Lett.} \textbf{\bibinfo{volume}{74}},
  \bibinfo{pages}{3029} (\bibinfo{year}{1995}).

\bibitem[{\citenamefont{Kolton et~al.}(1999)\citenamefont{Kolton,
  Dom\'{\i}nguez, and Gr\o{}nbech-Jensen}}]{Kolton99}
\bibinfo{author}{\bibfnamefont{A.~B.} \bibnamefont{Kolton}},
  \bibinfo{author}{\bibfnamefont{D.}~\bibnamefont{Dom\'{\i}nguez}},
  \bibnamefont{and}
  \bibinfo{author}{\bibfnamefont{N.}~\bibnamefont{Gr\o{}nbech-Jensen}},
  \bibinfo{journal}{Phys. Rev. Lett.} \textbf{\bibinfo{volume}{83}},
  \bibinfo{pages}{3061} (\bibinfo{year}{1999}).

\bibitem[{\citenamefont{Togawa et~al.}(2000)\citenamefont{Togawa, Abiru, Iwaya,
  Kitano, and Maeda}}]{Togawa00}
\bibinfo{author}{\bibfnamefont{Y.}~\bibnamefont{Togawa}},
  \bibinfo{author}{\bibfnamefont{R.}~\bibnamefont{Abiru}},
  \bibinfo{author}{\bibfnamefont{K.}~\bibnamefont{Iwaya}},
  \bibinfo{author}{\bibfnamefont{H.}~\bibnamefont{Kitano}}, \bibnamefont{and}
  \bibinfo{author}{\bibfnamefont{A.}~\bibnamefont{Maeda}},
  \bibinfo{journal}{Phys. Rev. Lett.} \textbf{\bibinfo{volume}{85}},
  \bibinfo{pages}{3716} (\bibinfo{year}{2000}).

\bibitem[{\citenamefont{Okuma et~al.}(2007)\citenamefont{Okuma, Inoue, and
  Kokubo}}]{Okuma07}
\bibinfo{author}{\bibfnamefont{S.}~\bibnamefont{Okuma}},
  \bibinfo{author}{\bibfnamefont{J.}~\bibnamefont{Inoue}}, \bibnamefont{and}
  \bibinfo{author}{\bibfnamefont{N.}~\bibnamefont{Kokubo}},
  \bibinfo{journal}{Phys. Rev. B} \textbf{\bibinfo{volume}{76}},
  \bibinfo{pages}{172503} (\bibinfo{year}{2007}).

\bibitem[{\citenamefont{D\'{\i}az et~al.}(2017)\citenamefont{D\'{\i}az,
  Reichhardt, Arovas, Saxena, and Reichhardt}}]{Diaz17}
\bibinfo{author}{\bibfnamefont{S.~A.} \bibnamefont{D\'{\i}az}},
  \bibinfo{author}{\bibfnamefont{C.~J.~O.} \bibnamefont{Reichhardt}},
  \bibinfo{author}{\bibfnamefont{D.~P.} \bibnamefont{Arovas}},
  \bibinfo{author}{\bibfnamefont{A.}~\bibnamefont{Saxena}}, \bibnamefont{and}
  \bibinfo{author}{\bibfnamefont{C.}~\bibnamefont{Reichhardt}},
  \bibinfo{journal}{Phys. Rev. B} \textbf{\bibinfo{volume}{96}},
  \bibinfo{pages}{085106} (\bibinfo{year}{2017}).

\bibitem[{\citenamefont{Sato et~al.}(2019)\citenamefont{Sato, Koshibae,
  Kikkawa, Yokouchi, Oike, Taguchi, Nagaosa, Tokura, and Kagawa}}]{Sato19}
\bibinfo{author}{\bibfnamefont{T.}~\bibnamefont{Sato}},
  \bibinfo{author}{\bibfnamefont{W.}~\bibnamefont{Koshibae}},
  \bibinfo{author}{\bibfnamefont{A.}~\bibnamefont{Kikkawa}},
  \bibinfo{author}{\bibfnamefont{T.}~\bibnamefont{Yokouchi}},
  \bibinfo{author}{\bibfnamefont{H.}~\bibnamefont{Oike}},
  \bibinfo{author}{\bibfnamefont{Y.}~\bibnamefont{Taguchi}},
  \bibinfo{author}{\bibfnamefont{N.}~\bibnamefont{Nagaosa}},
  \bibinfo{author}{\bibfnamefont{Y.}~\bibnamefont{Tokura}}, \bibnamefont{and}
  \bibinfo{author}{\bibfnamefont{F.}~\bibnamefont{Kagawa}},
  \bibinfo{journal}{Phys. Rev. B} \textbf{\bibinfo{volume}{100}},
  \bibinfo{pages}{094410} (\bibinfo{year}{2019}).

\bibitem[{\citenamefont{Sun et~al.}(2022)\citenamefont{Sun, Niu, Li, Liu,
  Pfeiffer, West, Wang, and Lin}}]{Sun22}
\bibinfo{author}{\bibfnamefont{J.}~\bibnamefont{Sun}},
  \bibinfo{author}{\bibfnamefont{J.}~\bibnamefont{Niu}},
  \bibinfo{author}{\bibfnamefont{Y.}~\bibnamefont{Li}},
  \bibinfo{author}{\bibfnamefont{Y.}~\bibnamefont{Liu}},
  \bibinfo{author}{\bibfnamefont{L.~N.} \bibnamefont{Pfeiffer}},
  \bibinfo{author}{\bibfnamefont{K.~W.} \bibnamefont{West}},
  \bibinfo{author}{\bibfnamefont{P.}~\bibnamefont{Wang}}, \bibnamefont{and}
  \bibinfo{author}{\bibfnamefont{X.}~\bibnamefont{Lin}},
  \bibinfo{journal}{Fund. Res.} \textbf{\bibinfo{volume}{2}},
  \bibinfo{pages}{178} (\bibinfo{year}{2022}).

\bibitem[{\citenamefont{Andrei et~al.}(1988)\citenamefont{Andrei, Deville,
  Glattli, Williams, Paris, and Etienne}}]{Andrei88}
\bibinfo{author}{\bibfnamefont{E.~Y.} \bibnamefont{Andrei}},
  \bibinfo{author}{\bibfnamefont{G.}~\bibnamefont{Deville}},
  \bibinfo{author}{\bibfnamefont{D.~C.} \bibnamefont{Glattli}},
  \bibinfo{author}{\bibfnamefont{F.~I.~B.} \bibnamefont{Williams}},
  \bibinfo{author}{\bibfnamefont{E.}~\bibnamefont{Paris}}, \bibnamefont{and}
  \bibinfo{author}{\bibfnamefont{B.}~\bibnamefont{Etienne}},
  \bibinfo{journal}{Phys. Rev. Lett.} \textbf{\bibinfo{volume}{60}},
  \bibinfo{pages}{2765} (\bibinfo{year}{1988}).

\bibitem[{\citenamefont{Goldman et~al.}(1990)\citenamefont{Goldman, Santos,
  Shayegan, and Cunningham}}]{Goldman90}
\bibinfo{author}{\bibfnamefont{V.~J.} \bibnamefont{Goldman}},
  \bibinfo{author}{\bibfnamefont{M.}~\bibnamefont{Santos}},
  \bibinfo{author}{\bibfnamefont{M.}~\bibnamefont{Shayegan}}, \bibnamefont{and}
  \bibinfo{author}{\bibfnamefont{J.~E.} \bibnamefont{Cunningham}},
  \bibinfo{journal}{Phys. Rev. Lett.} \textbf{\bibinfo{volume}{65}},
  \bibinfo{pages}{2189} (\bibinfo{year}{1990}).

\bibitem[{\citenamefont{Williams et~al.}(1991)\citenamefont{Williams, Wright,
  Clark, Andrei, Deville, Glattli, Probst, Etienne, Dorin, Foxon
  et~al.}}]{Williams91}
\bibinfo{author}{\bibfnamefont{F.~I.~B.} \bibnamefont{Williams}},
  \bibinfo{author}{\bibfnamefont{P.~A.} \bibnamefont{Wright}},
  \bibinfo{author}{\bibfnamefont{R.~G.} \bibnamefont{Clark}},
  \bibinfo{author}{\bibfnamefont{E.~Y.} \bibnamefont{Andrei}},
  \bibinfo{author}{\bibfnamefont{G.}~\bibnamefont{Deville}},
  \bibinfo{author}{\bibfnamefont{D.~C.} \bibnamefont{Glattli}},
  \bibinfo{author}{\bibfnamefont{O.}~\bibnamefont{Probst}},
  \bibinfo{author}{\bibfnamefont{B.}~\bibnamefont{Etienne}},
  \bibinfo{author}{\bibfnamefont{C.}~\bibnamefont{Dorin}},
  \bibinfo{author}{\bibfnamefont{C.~T.} \bibnamefont{Foxon}},
  \bibnamefont{et~al.}, \bibinfo{journal}{Phys. Rev. Lett.}
  \textbf{\bibinfo{volume}{66}}, \bibinfo{pages}{3285} (\bibinfo{year}{1991}).

\bibitem[{\citenamefont{Jiang et~al.}(1991)\citenamefont{Jiang, Stormer, Tsui,
  Pfeiffer, and West}}]{Jiang91}
\bibinfo{author}{\bibfnamefont{H.~W.} \bibnamefont{Jiang}},
  \bibinfo{author}{\bibfnamefont{H.~L.} \bibnamefont{Stormer}},
  \bibinfo{author}{\bibfnamefont{D.~C.} \bibnamefont{Tsui}},
  \bibinfo{author}{\bibfnamefont{L.~N.} \bibnamefont{Pfeiffer}},
  \bibnamefont{and} \bibinfo{author}{\bibfnamefont{K.~W.} \bibnamefont{West}},
  \bibinfo{journal}{Phys. Rev. B} \textbf{\bibinfo{volume}{44}},
  \bibinfo{pages}{8107} (\bibinfo{year}{1991}).

\bibitem[{\citenamefont{Zhu et~al.}(1994)\citenamefont{Zhu, Littlewood, and
  Millis}}]{Zhu94}
\bibinfo{author}{\bibfnamefont{X.}~\bibnamefont{Zhu}},
  \bibinfo{author}{\bibfnamefont{P.~B.} \bibnamefont{Littlewood}},
  \bibnamefont{and} \bibinfo{author}{\bibfnamefont{A.~J.}
  \bibnamefont{Millis}}, \bibinfo{journal}{Phys. Rev. B}
  \textbf{\bibinfo{volume}{50}}, \bibinfo{pages}{4600} (\bibinfo{year}{1994}).

\bibitem[{\citenamefont{Cha and Fertig}(1994{\natexlab{a}})}]{Cha94}
\bibinfo{author}{\bibfnamefont{M.-C.} \bibnamefont{Cha}} \bibnamefont{and}
  \bibinfo{author}{\bibfnamefont{H.~A.} \bibnamefont{Fertig}},
  \bibinfo{journal}{Phys. Rev. Lett.} \textbf{\bibinfo{volume}{73}},
  \bibinfo{pages}{870} (\bibinfo{year}{1994}{\natexlab{a}}).

\bibitem[{\citenamefont{Reichhardt et~al.}(2001)\citenamefont{Reichhardt,
  Olson, Gr\o{}nbech-Jensen, and Nori}}]{Reichhardt01}
\bibinfo{author}{\bibfnamefont{C.}~\bibnamefont{Reichhardt}},
  \bibinfo{author}{\bibfnamefont{C.~J.} \bibnamefont{Olson}},
  \bibinfo{author}{\bibfnamefont{N.}~\bibnamefont{Gr\o{}nbech-Jensen}},
  \bibnamefont{and} \bibinfo{author}{\bibfnamefont{F.}~\bibnamefont{Nori}},
  \bibinfo{journal}{Phys. Rev. Lett.} \textbf{\bibinfo{volume}{86}},
  \bibinfo{pages}{4354} (\bibinfo{year}{2001}).

\bibitem[{\citenamefont{Monceau}(2012)}]{Monceau12}
\bibinfo{author}{\bibfnamefont{P.}~\bibnamefont{Monceau}},
  \bibinfo{journal}{Adv. Phys.} \textbf{\bibinfo{volume}{61}},
  \bibinfo{pages}{325} (\bibinfo{year}{2012}).

\bibitem[{\citenamefont{Brussarski et~al.}(2018)\citenamefont{Brussarski, Li,
  Kravchenko, Shashkin, and Sarachik}}]{Brussarski18}
\bibinfo{author}{\bibfnamefont{P.}~\bibnamefont{Brussarski}},
  \bibinfo{author}{\bibfnamefont{S.}~\bibnamefont{Li}},
  \bibinfo{author}{\bibfnamefont{S.~V.} \bibnamefont{Kravchenko}},
  \bibinfo{author}{\bibfnamefont{A.~A.} \bibnamefont{Shashkin}},
  \bibnamefont{and} \bibinfo{author}{\bibfnamefont{M.~P.}
  \bibnamefont{Sarachik}}, \bibinfo{journal}{Nature Commun.}
  \textbf{\bibinfo{volume}{9}}, \bibinfo{pages}{3803} (\bibinfo{year}{2018}).

\bibitem[{\citenamefont{Yoon et~al.}(1999)\citenamefont{Yoon, Li, Shahar, Tsui,
  and Shayegan}}]{Yoon99}
\bibinfo{author}{\bibfnamefont{J.}~\bibnamefont{Yoon}},
  \bibinfo{author}{\bibfnamefont{C.~C.} \bibnamefont{Li}},
  \bibinfo{author}{\bibfnamefont{D.}~\bibnamefont{Shahar}},
  \bibinfo{author}{\bibfnamefont{D.~C.} \bibnamefont{Tsui}}, \bibnamefont{and}
  \bibinfo{author}{\bibfnamefont{M.}~\bibnamefont{Shayegan}},
  \bibinfo{journal}{Phys. Rev. Lett.} \textbf{\bibinfo{volume}{82}},
  \bibinfo{pages}{1744} (\bibinfo{year}{1999}).

\bibitem[{\citenamefont{Hossain et~al.}(2022)\citenamefont{Hossain, Ma,
  Villegas-Rosales, Chung, Pfeiffer, West, Baldwin, and Shayegan}}]{Hossain22}
\bibinfo{author}{\bibfnamefont{M.~S.} \bibnamefont{Hossain}},
  \bibinfo{author}{\bibfnamefont{M.~K.} \bibnamefont{Ma}},
  \bibinfo{author}{\bibfnamefont{K.~A.} \bibnamefont{Villegas-Rosales}},
  \bibinfo{author}{\bibfnamefont{Y.~J.} \bibnamefont{Chung}},
  \bibinfo{author}{\bibfnamefont{L.~N.} \bibnamefont{Pfeiffer}},
  \bibinfo{author}{\bibfnamefont{K.~W.} \bibnamefont{West}},
  \bibinfo{author}{\bibfnamefont{K.~W.} \bibnamefont{Baldwin}},
  \bibnamefont{and} \bibinfo{author}{\bibfnamefont{M.}~\bibnamefont{Shayegan}},
  \bibinfo{journal}{Phys. Rev. Lett.} \textbf{\bibinfo{volume}{129}},
  \bibinfo{pages}{036601} (\bibinfo{year}{2022}).

\bibitem[{\citenamefont{Rees et~al.}(2016)\citenamefont{Rees, Beysengulov, Lin,
  and Kono}}]{Rees16}
\bibinfo{author}{\bibfnamefont{D.~G.} \bibnamefont{Rees}},
  \bibinfo{author}{\bibfnamefont{N.~R.} \bibnamefont{Beysengulov}},
  \bibinfo{author}{\bibfnamefont{J.-J.} \bibnamefont{Lin}}, \bibnamefont{and}
  \bibinfo{author}{\bibfnamefont{K.}~\bibnamefont{Kono}},
  \bibinfo{journal}{Phys. Rev. Lett.} \textbf{\bibinfo{volume}{116}},
  \bibinfo{pages}{206801} (\bibinfo{year}{2016}).

\bibitem[{\citenamefont{Badrutdinov et~al.}(2016)\citenamefont{Badrutdinov,
  Smorodin, Rees, Lin, and Konstantinov}}]{Badrutdinov16}
\bibinfo{author}{\bibfnamefont{A.~O.} \bibnamefont{Badrutdinov}},
  \bibinfo{author}{\bibfnamefont{A.~V.} \bibnamefont{Smorodin}},
  \bibinfo{author}{\bibfnamefont{D.~G.} \bibnamefont{Rees}},
  \bibinfo{author}{\bibfnamefont{J.~Y.} \bibnamefont{Lin}}, \bibnamefont{and}
  \bibinfo{author}{\bibfnamefont{D.}~\bibnamefont{Konstantinov}},
  \bibinfo{journal}{Phys. Rev. B} \textbf{\bibinfo{volume}{94}},
  \bibinfo{pages}{195311} (\bibinfo{year}{2016}).

\bibitem[{\citenamefont{Lin et~al.}(2018)\citenamefont{Lin, Smorodin,
  Badrutdinov, and Konstantinov}}]{Lin18}
\bibinfo{author}{\bibfnamefont{J.-Y.} \bibnamefont{Lin}},
  \bibinfo{author}{\bibfnamefont{A.~V.} \bibnamefont{Smorodin}},
  \bibinfo{author}{\bibfnamefont{A.~O.} \bibnamefont{Badrutdinov}},
  \bibnamefont{and}
  \bibinfo{author}{\bibfnamefont{D.}~\bibnamefont{Konstantinov}},
  \bibinfo{journal}{Phys. Rev. B} \textbf{\bibinfo{volume}{98}},
  \bibinfo{pages}{085412} (\bibinfo{year}{2018}).

\bibitem[{\citenamefont{Fogler et~al.}(1996)\citenamefont{Fogler, Koulakov, and
  Shklovskii}}]{Fogler96}
\bibinfo{author}{\bibfnamefont{M.~M.} \bibnamefont{Fogler}},
  \bibinfo{author}{\bibfnamefont{A.~A.} \bibnamefont{Koulakov}},
  \bibnamefont{and} \bibinfo{author}{\bibfnamefont{B.~I.}
  \bibnamefont{Shklovskii}}, \bibinfo{journal}{Phys. Rev. B}
  \textbf{\bibinfo{volume}{54}}, \bibinfo{pages}{1853} (\bibinfo{year}{1996}).

\bibitem[{\citenamefont{Moessner and Chalker}(1996)}]{Moessner96}
\bibinfo{author}{\bibfnamefont{R.}~\bibnamefont{Moessner}} \bibnamefont{and}
  \bibinfo{author}{\bibfnamefont{J.~T.} \bibnamefont{Chalker}},
  \bibinfo{journal}{Phys. Rev. B} \textbf{\bibinfo{volume}{54}},
  \bibinfo{pages}{5006} (\bibinfo{year}{1996}).

\bibitem[{\citenamefont{Fradkin and Kivelson}(1999)}]{Fradkin99}
\bibinfo{author}{\bibfnamefont{E.}~\bibnamefont{Fradkin}} \bibnamefont{and}
  \bibinfo{author}{\bibfnamefont{S.~A.} \bibnamefont{Kivelson}},
  \bibinfo{journal}{Phys. Rev. B} \textbf{\bibinfo{volume}{59}},
  \bibinfo{pages}{8065} (\bibinfo{year}{1999}).

\bibitem[{\citenamefont{Lilly et~al.}(1999)\citenamefont{Lilly, Cooper,
  Eisenstein, Pfeiffer, and West}}]{Lilly99a}
\bibinfo{author}{\bibfnamefont{M.~P.} \bibnamefont{Lilly}},
  \bibinfo{author}{\bibfnamefont{K.~B.} \bibnamefont{Cooper}},
  \bibinfo{author}{\bibfnamefont{J.~P.} \bibnamefont{Eisenstein}},
  \bibinfo{author}{\bibfnamefont{L.~N.} \bibnamefont{Pfeiffer}},
  \bibnamefont{and} \bibinfo{author}{\bibfnamefont{K.~W.} \bibnamefont{West}},
  \bibinfo{journal}{Phys. Rev. Lett.} \textbf{\bibinfo{volume}{82}},
  \bibinfo{pages}{394} (\bibinfo{year}{1999}).

\bibitem[{\citenamefont{Cooper et~al.}(1999)\citenamefont{Cooper, Lilly,
  Eisenstein, Pfeiffer, and West}}]{Cooper99}
\bibinfo{author}{\bibfnamefont{K.~B.} \bibnamefont{Cooper}},
  \bibinfo{author}{\bibfnamefont{M.~P.} \bibnamefont{Lilly}},
  \bibinfo{author}{\bibfnamefont{J.~P.} \bibnamefont{Eisenstein}},
  \bibinfo{author}{\bibfnamefont{L.~N.} \bibnamefont{Pfeiffer}},
  \bibnamefont{and} \bibinfo{author}{\bibfnamefont{K.~W.} \bibnamefont{West}},
  \bibinfo{journal}{Phys. Rev. B} \textbf{\bibinfo{volume}{60}},
  \bibinfo{pages}{R11285} (\bibinfo{year}{1999}).

\bibitem[{\citenamefont{Reichhardt et~al.}(2003)\citenamefont{Reichhardt,
  Reichhardt, Martin, and Bishop}}]{Reichhardt03a}
\bibinfo{author}{\bibfnamefont{C.}~\bibnamefont{Reichhardt}},
  \bibinfo{author}{\bibfnamefont{C.~J.~O.} \bibnamefont{Reichhardt}},
  \bibinfo{author}{\bibfnamefont{I.}~\bibnamefont{Martin}}, \bibnamefont{and}
  \bibinfo{author}{\bibfnamefont{A.~R.} \bibnamefont{Bishop}},
  \bibinfo{journal}{Phys. Rev. Lett.} \textbf{\bibinfo{volume}{90}},
  \bibinfo{pages}{026401} (\bibinfo{year}{2003}).

\bibitem[{\citenamefont{Wang et~al.}(2015)\citenamefont{Wang, Fu, Du, Liu,
  Wang, Pfeiffer, West, Du, and Lin}}]{Wang15}
\bibinfo{author}{\bibfnamefont{X.}~\bibnamefont{Wang}},
  \bibinfo{author}{\bibfnamefont{H.}~\bibnamefont{Fu}},
  \bibinfo{author}{\bibfnamefont{L.}~\bibnamefont{Du}},
  \bibinfo{author}{\bibfnamefont{X.}~\bibnamefont{Liu}},
  \bibinfo{author}{\bibfnamefont{P.}~\bibnamefont{Wang}},
  \bibinfo{author}{\bibfnamefont{L.~N.} \bibnamefont{Pfeiffer}},
  \bibinfo{author}{\bibfnamefont{K.~W.} \bibnamefont{West}},
  \bibinfo{author}{\bibfnamefont{R.-R.} \bibnamefont{Du}}, \bibnamefont{and}
  \bibinfo{author}{\bibfnamefont{X.}~\bibnamefont{Lin}},
  \bibinfo{journal}{Phys. Rev. B} \textbf{\bibinfo{volume}{91}},
  \bibinfo{pages}{115301} (\bibinfo{year}{2015}).

\bibitem[{\citenamefont{Li et~al.}(2000)\citenamefont{Li, Yoon, Engel, Shahar,
  Tsui, and Shayegan}}]{Li00}
\bibinfo{author}{\bibfnamefont{C.-C.} \bibnamefont{Li}},
  \bibinfo{author}{\bibfnamefont{J.}~\bibnamefont{Yoon}},
  \bibinfo{author}{\bibfnamefont{L.~W.} \bibnamefont{Engel}},
  \bibinfo{author}{\bibfnamefont{D.}~\bibnamefont{Shahar}},
  \bibinfo{author}{\bibfnamefont{D.~C.} \bibnamefont{Tsui}}, \bibnamefont{and}
  \bibinfo{author}{\bibfnamefont{M.}~\bibnamefont{Shayegan}},
  \bibinfo{journal}{Phys. Rev. B} \textbf{\bibinfo{volume}{61}},
  \bibinfo{pages}{10905} (\bibinfo{year}{2000}).

\bibitem[{\citenamefont{Ye et~al.}(2002)\citenamefont{Ye, Engel, Tsui, Lewis,
  Pfeiffer, and West}}]{Ye02}
\bibinfo{author}{\bibfnamefont{P.~D.} \bibnamefont{Ye}},
  \bibinfo{author}{\bibfnamefont{L.~W.} \bibnamefont{Engel}},
  \bibinfo{author}{\bibfnamefont{D.~C.} \bibnamefont{Tsui}},
  \bibinfo{author}{\bibfnamefont{R.~M.} \bibnamefont{Lewis}},
  \bibinfo{author}{\bibfnamefont{L.~N.} \bibnamefont{Pfeiffer}},
  \bibnamefont{and} \bibinfo{author}{\bibfnamefont{K.}~\bibnamefont{West}},
  \bibinfo{journal}{Phys. Rev. Lett.} \textbf{\bibinfo{volume}{89}},
  \bibinfo{pages}{176802} (\bibinfo{year}{2002}).

\bibitem[{\citenamefont{Chen et~al.}(2004)\citenamefont{Chen, Lewis, Engel,
  Tsui, Ye, Wang, Pfeiffer, and West}}]{Chen04}
\bibinfo{author}{\bibfnamefont{Y.~P.} \bibnamefont{Chen}},
  \bibinfo{author}{\bibfnamefont{R.~M.} \bibnamefont{Lewis}},
  \bibinfo{author}{\bibfnamefont{L.~W.} \bibnamefont{Engel}},
  \bibinfo{author}{\bibfnamefont{D.~C.} \bibnamefont{Tsui}},
  \bibinfo{author}{\bibfnamefont{P.~D.} \bibnamefont{Ye}},
  \bibinfo{author}{\bibfnamefont{Z.~H.} \bibnamefont{Wang}},
  \bibinfo{author}{\bibfnamefont{L.~N.} \bibnamefont{Pfeiffer}},
  \bibnamefont{and} \bibinfo{author}{\bibfnamefont{K.~W.} \bibnamefont{West}},
  \bibinfo{journal}{Phys. Rev. Lett.} \textbf{\bibinfo{volume}{93}},
  \bibinfo{pages}{206805} (\bibinfo{year}{2004}).

\bibitem[{\citenamefont{Chen et~al.}(2006)\citenamefont{Chen, Sambandamurthy,
  Wang, Lewis, Engel, Tsui, Ye, Pfeiffer, and West}}]{Chen06}
\bibinfo{author}{\bibfnamefont{Y.~P.} \bibnamefont{Chen}},
  \bibinfo{author}{\bibfnamefont{G.}~\bibnamefont{Sambandamurthy}},
  \bibinfo{author}{\bibfnamefont{Z.~H.} \bibnamefont{Wang}},
  \bibinfo{author}{\bibfnamefont{R.~M.} \bibnamefont{Lewis}},
  \bibinfo{author}{\bibfnamefont{L.~W.} \bibnamefont{Engel}},
  \bibinfo{author}{\bibfnamefont{D.~C.} \bibnamefont{Tsui}},
  \bibinfo{author}{\bibfnamefont{P.~D.} \bibnamefont{Ye}},
  \bibinfo{author}{\bibfnamefont{L.~N.} \bibnamefont{Pfeiffer}},
  \bibnamefont{and} \bibinfo{author}{\bibfnamefont{K.~W.} \bibnamefont{West}},
  \bibinfo{journal}{Nat. Phys.} \textbf{\bibinfo{volume}{2}},
  \bibinfo{pages}{452} (\bibinfo{year}{2006}).

\bibitem[{\citenamefont{Jang et~al.}(2017)\citenamefont{Jang, Hunt, Pfeiffer,
  West, and Ashoori}}]{Jang17}
\bibinfo{author}{\bibfnamefont{J.}~\bibnamefont{Jang}},
  \bibinfo{author}{\bibfnamefont{B.~M.} \bibnamefont{Hunt}},
  \bibinfo{author}{\bibfnamefont{L.~N.} \bibnamefont{Pfeiffer}},
  \bibinfo{author}{\bibfnamefont{K.~W.} \bibnamefont{West}}, \bibnamefont{and}
  \bibinfo{author}{\bibfnamefont{R.~C.} \bibnamefont{Ashoori}},
  \bibinfo{journal}{Nature Phys.} \textbf{\bibinfo{volume}{13}},
  \bibinfo{pages}{340 } (\bibinfo{year}{2017}).

\bibitem[{\citenamefont{Zhang et~al.}(2015)\citenamefont{Zhang, Du, Manfra,
  Pfeiffer, and West}}]{Zhang15}
\bibinfo{author}{\bibfnamefont{C.}~\bibnamefont{Zhang}},
  \bibinfo{author}{\bibfnamefont{R.-R.} \bibnamefont{Du}},
  \bibinfo{author}{\bibfnamefont{M.~J.} \bibnamefont{Manfra}},
  \bibinfo{author}{\bibfnamefont{L.~N.} \bibnamefont{Pfeiffer}},
  \bibnamefont{and} \bibinfo{author}{\bibfnamefont{K.~W.} \bibnamefont{West}},
  \bibinfo{journal}{Phys. Rev. B} \textbf{\bibinfo{volume}{92}},
  \bibinfo{pages}{075434} (\bibinfo{year}{2015}).

\bibitem[{\citenamefont{Knighton et~al.}(2018)\citenamefont{Knighton, Wu,
  Huang, Serafin, Xia, Pfeiffer, and West}}]{Knighton18}
\bibinfo{author}{\bibfnamefont{T.}~\bibnamefont{Knighton}},
  \bibinfo{author}{\bibfnamefont{Z.}~\bibnamefont{Wu}},
  \bibinfo{author}{\bibfnamefont{J.}~\bibnamefont{Huang}},
  \bibinfo{author}{\bibfnamefont{A.}~\bibnamefont{Serafin}},
  \bibinfo{author}{\bibfnamefont{J.~S.} \bibnamefont{Xia}},
  \bibinfo{author}{\bibfnamefont{L.~N.} \bibnamefont{Pfeiffer}},
  \bibnamefont{and} \bibinfo{author}{\bibfnamefont{K.~W.} \bibnamefont{West}},
  \bibinfo{journal}{Phys. Rev. B} \textbf{\bibinfo{volume}{97}},
  \bibinfo{pages}{085135} (\bibinfo{year}{2018}).

\bibitem[{\citenamefont{Hossain et~al.}(2020)\citenamefont{Hossain, Ma,
  Villegas~Rosales, and Shayegan}}]{Hossain20}
\bibinfo{author}{\bibfnamefont{M.~S.} \bibnamefont{Hossain}},
  \bibinfo{author}{\bibfnamefont{M.~K.} \bibnamefont{Ma}},
  \bibinfo{author}{\bibfnamefont{K.~A.} \bibnamefont{Villegas~Rosales}},
  \bibnamefont{and} \bibinfo{author}{\bibfnamefont{M.}~\bibnamefont{Shayegan}},
  \bibinfo{journal}{Proc. Natl. Acad. Sci. (USA)}
  \textbf{\bibinfo{volume}{117}}, \bibinfo{pages}{32244}
  (\bibinfo{year}{2020}).

\bibitem[{\citenamefont{Shingla et~al.}(2021)\citenamefont{Shingla, Myers,
  Pfeiffer, Baldwin, and Cs{\' a}thy}}]{Shingla21}
\bibinfo{author}{\bibfnamefont{V.}~\bibnamefont{Shingla}},
  \bibinfo{author}{\bibfnamefont{S.~A.} \bibnamefont{Myers}},
  \bibinfo{author}{\bibfnamefont{L.~N.} \bibnamefont{Pfeiffer}},
  \bibinfo{author}{\bibfnamefont{K.~W.} \bibnamefont{Baldwin}},
  \bibnamefont{and} \bibinfo{author}{\bibfnamefont{G.~A.} \bibnamefont{Cs{\'
  a}thy}}, \bibinfo{journal}{Commun. Phys.} \textbf{\bibinfo{volume}{4}},
  \bibinfo{pages}{204} (\bibinfo{year}{2021}).

\bibitem[{\citenamefont{Jo et~al.}(2018)\citenamefont{Jo, Deng, Liu, Pfeiffer,
  West, Baldwin, and Shayegan}}]{Jo18}
\bibinfo{author}{\bibfnamefont{I.}~\bibnamefont{Jo}},
  \bibinfo{author}{\bibfnamefont{H.}~\bibnamefont{Deng}},
  \bibinfo{author}{\bibfnamefont{Y.}~\bibnamefont{Liu}},
  \bibinfo{author}{\bibfnamefont{L.~N.} \bibnamefont{Pfeiffer}},
  \bibinfo{author}{\bibfnamefont{K.~W.} \bibnamefont{West}},
  \bibinfo{author}{\bibfnamefont{K.~W.} \bibnamefont{Baldwin}},
  \bibnamefont{and} \bibinfo{author}{\bibfnamefont{M.}~\bibnamefont{Shayegan}},
  \bibinfo{journal}{Phys. Rev. Lett.} \textbf{\bibinfo{volume}{120}},
  \bibinfo{pages}{016802} (\bibinfo{year}{2018}).

\bibitem[{\citenamefont{Cs\'athy et~al.}(2007)\citenamefont{Cs\'athy, Tsui,
  Pfeiffer, and West}}]{Csathy07}
\bibinfo{author}{\bibfnamefont{G.~A.} \bibnamefont{Cs\'athy}},
  \bibinfo{author}{\bibfnamefont{D.~C.} \bibnamefont{Tsui}},
  \bibinfo{author}{\bibfnamefont{L.~N.} \bibnamefont{Pfeiffer}},
  \bibnamefont{and} \bibinfo{author}{\bibfnamefont{K.~W.} \bibnamefont{West}},
  \bibinfo{journal}{Phys. Rev. Lett.} \textbf{\bibinfo{volume}{98}},
  \bibinfo{pages}{066805} (\bibinfo{year}{2007}).

\bibitem[{\citenamefont{Madathil et~al.}(2022)\citenamefont{Madathil,
  Villegas~Rosales, Wang, Chung, Pfeiffer, Baldwin, West, and
  Shayegan}}]{Madathil22}
\bibinfo{author}{\bibfnamefont{P.~T.} \bibnamefont{Madathil}},
  \bibinfo{author}{\bibfnamefont{K.~A.} \bibnamefont{Villegas~Rosales}},
  \bibinfo{author}{\bibfnamefont{C.}~\bibnamefont{Wang}},
  \bibinfo{author}{\bibfnamefont{E.~Y.} \bibnamefont{Chung}},
  \bibinfo{author}{\bibfnamefont{L.~N.} \bibnamefont{Pfeiffer}},
  \bibinfo{author}{\bibfnamefont{K.}~\bibnamefont{Baldwin}},
  \bibinfo{author}{\bibfnamefont{K.~W.} \bibnamefont{West}}, \bibnamefont{and}
  \bibinfo{author}{\bibfnamefont{M.}~\bibnamefont{Shayegan}},
  \bibinfo{journal}{Bull. Am. Phys. Soc.} \textbf{\bibinfo{volume}{67}},
  \bibinfo{pages}{1141} (\bibinfo{year}{2022}).

\bibitem[{\citenamefont{Jaroszy\ifmmode~\acute{n}\else \'{n}\fi{}ski
  et~al.}(2002)\citenamefont{Jaroszy\ifmmode~\acute{n}\else \'{n}\fi{}ski,
  Popovi\ifmmode~\acute{c}\else \'{c}\fi{}, and Klapwijk}}]{Jaroszynski02}
\bibinfo{author}{\bibfnamefont{J.}~\bibnamefont{Jaroszy\ifmmode~\acute{n}\else
  \'{n}\fi{}ski}},
  \bibinfo{author}{\bibfnamefont{D.}~\bibnamefont{Popovi\ifmmode~\acute{c}\else
  \'{c}\fi{}}}, \bibnamefont{and} \bibinfo{author}{\bibfnamefont{T.~M.}
  \bibnamefont{Klapwijk}}, \bibinfo{journal}{Phys. Rev. Lett.}
  \textbf{\bibinfo{volume}{89}}, \bibinfo{pages}{276401}
  (\bibinfo{year}{2002}).

\bibitem[{\citenamefont{Leturcq et~al.}(2003)\citenamefont{Leturcq, L'H\^ote,
  Tourbot, Mellor, and Henini}}]{Leturcq03}
\bibinfo{author}{\bibfnamefont{R.}~\bibnamefont{Leturcq}},
  \bibinfo{author}{\bibfnamefont{D.}~\bibnamefont{L'H\^ote}},
  \bibinfo{author}{\bibfnamefont{R.}~\bibnamefont{Tourbot}},
  \bibinfo{author}{\bibfnamefont{C.~J.} \bibnamefont{Mellor}},
  \bibnamefont{and} \bibinfo{author}{\bibfnamefont{M.}~\bibnamefont{Henini}},
  \bibinfo{journal}{Phys. Rev. Lett.} \textbf{\bibinfo{volume}{90}},
  \bibinfo{pages}{076402} (\bibinfo{year}{2003}).

\bibitem[{\citenamefont{Jaroszy\ifmmode~\acute{n}\else \'{n}\fi{}ski
  et~al.}(2004)\citenamefont{Jaroszy\ifmmode~\acute{n}\else \'{n}\fi{}ski,
  Popovi\ifmmode~\acute{c}\else \'{c}\fi{}, and Klapwijk}}]{Jaroszynski04}
\bibinfo{author}{\bibfnamefont{J.}~\bibnamefont{Jaroszy\ifmmode~\acute{n}\else
  \'{n}\fi{}ski}},
  \bibinfo{author}{\bibfnamefont{D.}~\bibnamefont{Popovi\ifmmode~\acute{c}\else
  \'{c}\fi{}}}, \bibnamefont{and} \bibinfo{author}{\bibfnamefont{T.~M.}
  \bibnamefont{Klapwijk}}, \bibinfo{journal}{Phys. Rev. Lett.}
  \textbf{\bibinfo{volume}{92}}, \bibinfo{pages}{226403}
  (\bibinfo{year}{2004}).

\bibitem[{\citenamefont{Abrahams et~al.}(2001)\citenamefont{Abrahams,
  Kravchenko, and Sarachik}}]{Abrahams01}
\bibinfo{author}{\bibfnamefont{E.}~\bibnamefont{Abrahams}},
  \bibinfo{author}{\bibfnamefont{S.~V.} \bibnamefont{Kravchenko}},
  \bibnamefont{and} \bibinfo{author}{\bibfnamefont{M.~P.}
  \bibnamefont{Sarachik}}, \bibinfo{journal}{Rev. Mod. Phys.}
  \textbf{\bibinfo{volume}{73}}, \bibinfo{pages}{251} (\bibinfo{year}{2001}).

\bibitem[{\citenamefont{Spivak et~al.}(2010)\citenamefont{Spivak, Kravchenko,
  Kivelson, and Gao}}]{Spivak10}
\bibinfo{author}{\bibfnamefont{B.}~\bibnamefont{Spivak}},
  \bibinfo{author}{\bibfnamefont{S.~V.} \bibnamefont{Kravchenko}},
  \bibinfo{author}{\bibfnamefont{S.~A.} \bibnamefont{Kivelson}},
  \bibnamefont{and} \bibinfo{author}{\bibfnamefont{X.~P.~A.}
  \bibnamefont{Gao}}, \bibinfo{journal}{Rev. Mod. Phys.}
  \textbf{\bibinfo{volume}{82}}, \bibinfo{pages}{1743} (\bibinfo{year}{2010}).

\bibitem[{\citenamefont{Reichhardt and Olson~Reichhardt}(2004)}]{Reichhardt04}
\bibinfo{author}{\bibfnamefont{C.}~\bibnamefont{Reichhardt}} \bibnamefont{and}
  \bibinfo{author}{\bibfnamefont{C.~J.} \bibnamefont{Olson~Reichhardt}},
  \bibinfo{journal}{Phys. Rev. Lett.} \textbf{\bibinfo{volume}{93}},
  \bibinfo{pages}{176405} (\bibinfo{year}{2004}).

\bibitem[{\citenamefont{Smole{\' n}ski et~al.}(2021)\citenamefont{Smole{\'
  n}ski, Dolgirev, Kuhlenkamp, Popert, Shimazaki, Back, Lu, Kroner, Watanabe,
  Taniguchi et~al.}}]{Smolenski21}
\bibinfo{author}{\bibfnamefont{T.}~\bibnamefont{Smole{\' n}ski}},
  \bibinfo{author}{\bibfnamefont{P.~E.} \bibnamefont{Dolgirev}},
  \bibinfo{author}{\bibfnamefont{C.}~\bibnamefont{Kuhlenkamp}},
  \bibinfo{author}{\bibfnamefont{A.}~\bibnamefont{Popert}},
  \bibinfo{author}{\bibfnamefont{Y.}~\bibnamefont{Shimazaki}},
  \bibinfo{author}{\bibfnamefont{P.}~\bibnamefont{Back}},
  \bibinfo{author}{\bibfnamefont{X.}~\bibnamefont{Lu}},
  \bibinfo{author}{\bibfnamefont{M.}~\bibnamefont{Kroner}},
  \bibinfo{author}{\bibfnamefont{K.}~\bibnamefont{Watanabe}},
  \bibinfo{author}{\bibfnamefont{T.}~\bibnamefont{Taniguchi}},
  \bibnamefont{et~al.}, \bibinfo{journal}{Nature (London)}
  \textbf{\bibinfo{volume}{595}}, \bibinfo{pages}{53} (\bibinfo{year}{2021}).

\bibitem[{\citenamefont{Li et~al.}(2021)\citenamefont{Li, Li, Regan, Wang,
  Zhao, Kahn, Yumigeta, Blei, Taniguchi, Watanabe et~al.}}]{Li21}
\bibinfo{author}{\bibfnamefont{H.}~\bibnamefont{Li}},
  \bibinfo{author}{\bibfnamefont{S.}~\bibnamefont{Li}},
  \bibinfo{author}{\bibfnamefont{E.~C.} \bibnamefont{Regan}},
  \bibinfo{author}{\bibfnamefont{D.}~\bibnamefont{Wang}},
  \bibinfo{author}{\bibfnamefont{W.}~\bibnamefont{Zhao}},
  \bibinfo{author}{\bibfnamefont{S.}~\bibnamefont{Kahn}},
  \bibinfo{author}{\bibfnamefont{K.}~\bibnamefont{Yumigeta}},
  \bibinfo{author}{\bibfnamefont{M.}~\bibnamefont{Blei}},
  \bibinfo{author}{\bibfnamefont{T.}~\bibnamefont{Taniguchi}},
  \bibinfo{author}{\bibfnamefont{K.}~\bibnamefont{Watanabe}},
  \bibnamefont{et~al.}, \bibinfo{journal}{Nature (London)}
  \textbf{\bibinfo{volume}{597}}, \bibinfo{pages}{650} (\bibinfo{year}{2021}).

\bibitem[{\citenamefont{Padhi et~al.}(2018)\citenamefont{Padhi, Setty, and
  Phillips}}]{Padhi18}
\bibinfo{author}{\bibfnamefont{B.}~\bibnamefont{Padhi}},
  \bibinfo{author}{\bibfnamefont{C.}~\bibnamefont{Setty}}, \bibnamefont{and}
  \bibinfo{author}{\bibfnamefont{P.~W.} \bibnamefont{Phillips}},
  \bibinfo{journal}{Nano Lett.} \textbf{\bibinfo{volume}{18}},
  \bibinfo{pages}{6175} (\bibinfo{year}{2018}).

\bibitem[{\citenamefont{Zhou et~al.}(2021)\citenamefont{Zhou, Sung, Brutschea,
  Esterlis, Wang, Scuri, Gelly, Heo, Taniguchi, Watanabe et~al.}}]{Zhou21}
\bibinfo{author}{\bibfnamefont{Y.}~\bibnamefont{Zhou}},
  \bibinfo{author}{\bibfnamefont{J.}~\bibnamefont{Sung}},
  \bibinfo{author}{\bibfnamefont{E.}~\bibnamefont{Brutschea}},
  \bibinfo{author}{\bibfnamefont{I.}~\bibnamefont{Esterlis}},
  \bibinfo{author}{\bibfnamefont{Y.}~\bibnamefont{Wang}},
  \bibinfo{author}{\bibfnamefont{G.}~\bibnamefont{Scuri}},
  \bibinfo{author}{\bibfnamefont{R.~J.} \bibnamefont{Gelly}},
  \bibinfo{author}{\bibfnamefont{H.}~\bibnamefont{Heo}},
  \bibinfo{author}{\bibfnamefont{T.}~\bibnamefont{Taniguchi}},
  \bibinfo{author}{\bibfnamefont{K.}~\bibnamefont{Watanabe}},
  \bibnamefont{et~al.}, \bibinfo{journal}{Nature (London)}
  \textbf{\bibinfo{volume}{595}}, \bibinfo{pages}{48} (\bibinfo{year}{2021}).

\bibitem[{\citenamefont{Padhi et~al.}(2021)\citenamefont{Padhi, Chitra, and
  Phillips}}]{Padhi21}
\bibinfo{author}{\bibfnamefont{B.}~\bibnamefont{Padhi}},
  \bibinfo{author}{\bibfnamefont{R.}~\bibnamefont{Chitra}}, \bibnamefont{and}
  \bibinfo{author}{\bibfnamefont{P.~W.} \bibnamefont{Phillips}},
  \bibinfo{journal}{Phys. Rev. B} \textbf{\bibinfo{volume}{103}},
  \bibinfo{pages}{125146} (\bibinfo{year}{2021}).

\bibitem[{\citenamefont{Chung et~al.}(2021)\citenamefont{Chung,
  Villegas~Rosales, Baldwin, Madathil, West, Shayegan, and Pfeiffer}}]{Chung21}
\bibinfo{author}{\bibfnamefont{Y.~J.} \bibnamefont{Chung}},
  \bibinfo{author}{\bibfnamefont{K.~A.} \bibnamefont{Villegas~Rosales}},
  \bibinfo{author}{\bibfnamefont{K.~W.} \bibnamefont{Baldwin}},
  \bibinfo{author}{\bibfnamefont{P.~T.} \bibnamefont{Madathil}},
  \bibinfo{author}{\bibfnamefont{K.~W.} \bibnamefont{West}},
  \bibinfo{author}{\bibfnamefont{M.}~\bibnamefont{Shayegan}}, \bibnamefont{and}
  \bibinfo{author}{\bibfnamefont{L.~N.} \bibnamefont{Pfeiffer}},
  \bibinfo{journal}{Nature Mater.} \textbf{\bibinfo{volume}{20}},
  \bibinfo{pages}{632} (\bibinfo{year}{2021}).

\bibitem[{\citenamefont{Shayegan}(2022)}]{Shayegan22}
\bibinfo{author}{\bibfnamefont{M.}~\bibnamefont{Shayegan}},
  \bibinfo{journal}{Nature Rev. Phys.} \textbf{\bibinfo{volume}{4}},
  \bibinfo{pages}{212} (\bibinfo{year}{2022}).

\bibitem[{\citenamefont{Cha and Fertig}(1994{\natexlab{b}})}]{Cha94a}
\bibinfo{author}{\bibfnamefont{M.-C.} \bibnamefont{Cha}} \bibnamefont{and}
  \bibinfo{author}{\bibfnamefont{H.~A.} \bibnamefont{Fertig}},
  \bibinfo{journal}{Phys. Rev. Lett.} \textbf{\bibinfo{volume}{73}},
  \bibinfo{pages}{870} (\bibinfo{year}{1994}{\natexlab{b}}).

\bibitem[{\citenamefont{Reichhardt and Reichhardt}(2021)}]{Reichhardt21}
\bibinfo{author}{\bibfnamefont{C.}~\bibnamefont{Reichhardt}} \bibnamefont{and}
  \bibinfo{author}{\bibfnamefont{C.~J.~O.} \bibnamefont{Reichhardt}},
  \bibinfo{journal}{Phys. Rev. B} \textbf{\bibinfo{volume}{103}},
  \bibinfo{pages}{125107} (\bibinfo{year}{2021}).

\bibitem[{\citenamefont{Reichhardt et~al.}(2022)\citenamefont{Reichhardt,
  Reichhardt, and Milosevic}}]{Reichhardt22}
\bibinfo{author}{\bibfnamefont{C.}~\bibnamefont{Reichhardt}},
  \bibinfo{author}{\bibfnamefont{C.~J.~O.} \bibnamefont{Reichhardt}},
  \bibnamefont{and}
  \bibinfo{author}{\bibfnamefont{M.}~\bibnamefont{Milosevic}},
  \bibinfo{journal}{Rev. Mod. Phys.} p. \bibinfo{pages}{in press}
  (\bibinfo{year}{2022}).

\bibitem[{\citenamefont{Kolton et~al.}(2001)\citenamefont{Kolton,
  Dom\'{\i}nguez, and Gr\o{}nbech-Jensen}}]{Kolton01}
\bibinfo{author}{\bibfnamefont{A.~B.} \bibnamefont{Kolton}},
  \bibinfo{author}{\bibfnamefont{D.}~\bibnamefont{Dom\'{\i}nguez}},
  \bibnamefont{and}
  \bibinfo{author}{\bibfnamefont{N.}~\bibnamefont{Gr\o{}nbech-Jensen}},
  \bibinfo{journal}{Phys. Rev. Lett.} \textbf{\bibinfo{volume}{86}},
  \bibinfo{pages}{4112} (\bibinfo{year}{2001}).

\bibitem[{\citenamefont{Okuma et~al.}(2009)\citenamefont{Okuma, Yamazaki, and
  Kokubo}}]{Okuma09}
\bibinfo{author}{\bibfnamefont{S.}~\bibnamefont{Okuma}},
  \bibinfo{author}{\bibfnamefont{Y.}~\bibnamefont{Yamazaki}}, \bibnamefont{and}
  \bibinfo{author}{\bibfnamefont{N.}~\bibnamefont{Kokubo}},
  \bibinfo{journal}{Phys. Rev. B} \textbf{\bibinfo{volume}{80}},
  \bibinfo{pages}{220501} (\bibinfo{year}{2009}).

\bibitem[{\citenamefont{Sato et~al.}(2020)\citenamefont{Sato, Kikkawa, Taguchi,
  Tokura, and Kagawa}}]{Sato20}
\bibinfo{author}{\bibfnamefont{T.}~\bibnamefont{Sato}},
  \bibinfo{author}{\bibfnamefont{A.}~\bibnamefont{Kikkawa}},
  \bibinfo{author}{\bibfnamefont{Y.}~\bibnamefont{Taguchi}},
  \bibinfo{author}{\bibfnamefont{Y.}~\bibnamefont{Tokura}}, \bibnamefont{and}
  \bibinfo{author}{\bibfnamefont{F.}~\bibnamefont{Kagawa}},
  \bibinfo{journal}{Phys. Rev. B} \textbf{\bibinfo{volume}{102}},
  \bibinfo{pages}{180411} (\bibinfo{year}{2020}).

\bibitem[{\citenamefont{Bennaceur et~al.}(2018)\citenamefont{Bennaceur, Lupien,
  Reulet, Gervais, Pfeiffer, and West}}]{Bennaceur18}
\bibinfo{author}{\bibfnamefont{K.}~\bibnamefont{Bennaceur}},
  \bibinfo{author}{\bibfnamefont{C.}~\bibnamefont{Lupien}},
  \bibinfo{author}{\bibfnamefont{B.}~\bibnamefont{Reulet}},
  \bibinfo{author}{\bibfnamefont{G.}~\bibnamefont{Gervais}},
  \bibinfo{author}{\bibfnamefont{L.~N.} \bibnamefont{Pfeiffer}},
  \bibnamefont{and} \bibinfo{author}{\bibfnamefont{K.~W.} \bibnamefont{West}},
  \bibinfo{journal}{Phys. Rev. Lett.} \textbf{\bibinfo{volume}{120}},
  \bibinfo{pages}{136801} (\bibinfo{year}{2018}).

\bibitem[{\citenamefont{Palassini and Goethe}(2012)}]{Palassini12}
\bibinfo{author}{\bibfnamefont{M.}~\bibnamefont{Palassini}} \bibnamefont{and}
  \bibinfo{author}{\bibfnamefont{M.}~\bibnamefont{Goethe}},
  \bibinfo{journal}{J. Phys.: Conf. Ser} \textbf{\bibinfo{volume}{376}},
  \bibinfo{pages}{012009} (\bibinfo{year}{2012}).

\bibitem[{\citenamefont{Olson et~al.}(1997)\citenamefont{Olson, Reichhardt, and
  Nori}}]{Olson97}
\bibinfo{author}{\bibfnamefont{C.~J.} \bibnamefont{Olson}},
  \bibinfo{author}{\bibfnamefont{C.}~\bibnamefont{Reichhardt}},
  \bibnamefont{and} \bibinfo{author}{\bibfnamefont{F.}~\bibnamefont{Nori}},
  \bibinfo{journal}{Phys. Rev. B} \textbf{\bibinfo{volume}{56}},
  \bibinfo{pages}{6175} (\bibinfo{year}{1997}).

\bibitem[{\citenamefont{Bassler and Paczuski}(1998)}]{Bassler98}
\bibinfo{author}{\bibfnamefont{K.~E.} \bibnamefont{Bassler}} \bibnamefont{and}
  \bibinfo{author}{\bibfnamefont{M.}~\bibnamefont{Paczuski}},
  \bibinfo{journal}{Phys. Rev. Lett.} \textbf{\bibinfo{volume}{81}},
  \bibinfo{pages}{3761} (\bibinfo{year}{1998}).

\bibitem[{\citenamefont{D\'{\i}az et~al.}(2018)\citenamefont{D\'{\i}az,
  Reichhardt, Arovas, Saxena, and Reichhardt}}]{Diaz18}
\bibinfo{author}{\bibfnamefont{S.~A.} \bibnamefont{D\'{\i}az}},
  \bibinfo{author}{\bibfnamefont{C.}~\bibnamefont{Reichhardt}},
  \bibinfo{author}{\bibfnamefont{D.~P.} \bibnamefont{Arovas}},
  \bibinfo{author}{\bibfnamefont{A.}~\bibnamefont{Saxena}}, \bibnamefont{and}
  \bibinfo{author}{\bibfnamefont{C.~J.~O.} \bibnamefont{Reichhardt}},
  \bibinfo{journal}{Phys. Rev. Lett.} \textbf{\bibinfo{volume}{120}},
  \bibinfo{pages}{117203} (\bibinfo{year}{2018}).

\bibitem[{\citenamefont{Baity et~al.}(2018)\citenamefont{Baity, Sasagawa, and
  Popovi\'{c}}}]{Baity18}
\bibinfo{author}{\bibfnamefont{P.~G.} \bibnamefont{Baity}},
  \bibinfo{author}{\bibfnamefont{T.}~\bibnamefont{Sasagawa}}, \bibnamefont{and}
  \bibinfo{author}{\bibfnamefont{D.}~\bibnamefont{Popovi\'{c}}},
  \bibinfo{journal}{Phys. Rev. Lett.} \textbf{\bibinfo{volume}{120}},
  \bibinfo{pages}{156602} (\bibinfo{year}{2018}).

\bibitem[{\citenamefont{Perkovi\'{c} et~al.}(1995)\citenamefont{Perkovi\'{c},
  Dahmen, and Sethna}}]{Perkovic95}
\bibinfo{author}{\bibfnamefont{O.}~\bibnamefont{Perkovi\'{c}}},
  \bibinfo{author}{\bibfnamefont{K.}~\bibnamefont{Dahmen}}, \bibnamefont{and}
  \bibinfo{author}{\bibfnamefont{J.~P.} \bibnamefont{Sethna}},
  \bibinfo{journal}{Phys. Rev. Lett.} \textbf{\bibinfo{volume}{75}},
  \bibinfo{pages}{4528} (\bibinfo{year}{1995}).

\bibitem[{\citenamefont{Reichhardt and Reichhardt}(2009)}]{Reichhardt09}
\bibinfo{author}{\bibfnamefont{C.}~\bibnamefont{Reichhardt}} \bibnamefont{and}
  \bibinfo{author}{\bibfnamefont{C.~J.~O.} \bibnamefont{Reichhardt}},
  \bibinfo{journal}{Phys. Rev. Lett.} \textbf{\bibinfo{volume}{103}},
  \bibinfo{pages}{168301} (\bibinfo{year}{2009}).

\bibitem[{\citenamefont{Kaji et~al.}(2022)\citenamefont{Kaji, Maegochi, Ienaga,
  Kaneko, and Okuma}}]{Kaji22}
\bibinfo{author}{\bibfnamefont{T.}~\bibnamefont{Kaji}},
  \bibinfo{author}{\bibfnamefont{S.}~\bibnamefont{Maegochi}},
  \bibinfo{author}{\bibfnamefont{K.}~\bibnamefont{Ienaga}},
  \bibinfo{author}{\bibfnamefont{S.}~\bibnamefont{Kaneko}}, \bibnamefont{and}
  \bibinfo{author}{\bibfnamefont{S.}~\bibnamefont{Okuma}},
  \bibinfo{journal}{Sci. Rep.} \textbf{\bibinfo{volume}{12}},
  \bibinfo{pages}{1542} (\bibinfo{year}{2022}).

\end{thebibliography}

\end{document}